\documentclass[12pt]{iopart}
\usepackage{graphicx}
\usepackage{iopams}
\eqnobysec

\begin{document}
\title{Zero temperature correlation functions for the impenetrable fermion gas}

\author{Vadim V Cheianov\dag\ and M B Zvonarev\ddag\S}

\address{\dag\ NORDITA, Blegdamsvej 17,  Copenhagen {\O}  DK 2100, Denmark}

\address{\ddag\ \O rsted Laboratory, Niels Bohr Institute for APG, Universitetsparken 5, Copenhagen {\O} DK 2100, Denmark}

\address{\S\ Petersburg Department of Steklov Institute of Mathematics,
Fontanka 27, St~Petersburg 191023, Russia}

\eads{\mailto{cheianov@nordita.dk}, \mailto{zvonarev@fys.ku.dk}}

\begin{abstract}
We calculate the long time and distance asymptotics of the
one-particle correlation functions in the model of impenetrable
spin $1/2$ fermions in 1+1 dimensions. We consider the spin
disordered zero temperature regime, which occurs when the limit
$T\to0$ is taken at a positive chemical potential. The asymptotic
expressions are found from the asymptotic solution of the matrix
Riemann-Hilbert problem related to the determinant representation
of the correlation functions.
\end{abstract}


\submitto{\JPA}



\section{Introduction}
Consider the one-dimensional  gas of spin $1/2$ fermions
interacting via the $\delta$-potential. The Hamiltonian of this
model is
\begin{equation}
H=\int\rmd x\left[-\sum_{\alpha=\uparrow,\downarrow}
\psi_{\alpha}^{\dagger}(x)\partial_{x}^2\psi_{\alpha}(x)
+Un_{\uparrow}(x)n_{\downarrow}(x) - \mu n(x) \right]. \label{H}
\end{equation}
The fermionic fields $\psi_{\alpha}$ ($\alpha$ is a spin index,
$\alpha=\uparrow,\downarrow$) satisfy canonical equal-time
anti-commutational relations
\begin{equation}
\psi_{\alpha}(x)\psi_{\beta}^{\dagger}(y)
+\psi_{\beta}^{\dagger}(y)\psi_{\alpha}(x)
=\delta_{\alpha\beta}\delta(x-y).
\label{canonical}
\end{equation}
The operators $n_{\alpha}(x)=\psi^\dagger_{\alpha}(x)
\psi_{\alpha}(x)$ are the density operators for spin-up
(spin-down) fermions and $n(x)=n_\uparrow(x)+n_\downarrow(x)$ is
the total fermion density operator. In this paper we will consider
the infinite $U$ limit of one-particle correlation functions
\begin{equation}
G_h(x,t)=\langle \psi^\dagger_{\uparrow}(x,t)
\psi_{\uparrow}(0,0)\rangle \label{Gholedef}
\end{equation}
and
\begin{eqnarray}
G_e(x,t)=\langle \psi_{\uparrow}(x,t)
\psi^\dagger_{\uparrow}(0,0)\rangle \label{Gelectrondef}
\end{eqnarray}
where the average $\langle\rangle$ is taken over the thermal
ensemble at a given temperature $T$ and chemical potential $\mu.$
We will calculate the large $x$ and $t$ asymptotics of the
correlation functions \eref{Gholedef} and \eref{Gelectrondef}
assuming that (a) the limit of infinite repulsion $U=\infty$ is
taken at finite positive chemical potential $\mu,$ and (b), the
zero temperature limit is taken afterwards. The ground state of
the system is infinitely degenerate at infinite $U.$ This
degeneracy leads to a non-trivial (non-conformal) behaviour of the
correlation functions \eref{Gholedef} and \eref{Gelectrondef} at
zero temperature. The physics of this phenomenon were presented
earlier \cite{CZ-03}, here we present a detailed derivation of
these results.

Our starting point is the determinant representation for
$G_h(x,t)$ and $G_e(x,t),$  recently derived by Izergin and Pronko
\cite{IP-98}. We relate the objects entering this representation
to the solution of the corresponding matrix Riemann-Hilbert
problem (RH problem) \cite{GIKP-98}. We perform the asymptotic
analysis of this RH problem making use of previous work
\cite{DKMVZ-99,D-book}. The large $x$ asymptotics of the
correlation functions at $t=0$ are given in subsection
\ref{answer0}. The  asymptotics of the time-dependent correlation
functions in the space-like region are given in subsection
\ref{answerx}. The asymptotics in the time-like region for
$x/t=\mathrm{const}>0$ are given in subsection \ref{answert}. The
large $t$ asymptotics of the correlation functions for $x=0$ are
given in section \ref{ks=0}.

\section{Determinant representation and the Riemann-Hilbert problem
for the correlation functions of the model \label{DetRep} }

This section is introductory. In subsection \ref{elrel} we recall
some elementary relations for the correlation functions
\eref{Gholedef} and \eref{Gelectrondef}. In subsection
\ref{detrep} we recall the results of \cite{IP-98} on the
determinant representation for these correlation functions. In
subsection \ref{residue} we examine the analytic properties of the
functions entering this determinant representation. In subsection
\ref{RHsection}, following \cite{GIKP-98}, we formulate the matrix
RH problem related to the determinant representation of the
correlation functions \eref{Gholedef} and \eref{Gelectrondef}.

\subsection{Elementary relations \label{elrel}}
In this subsection we recall some elementary relations for the
correlation functions \eref{Gholedef} and \eref{Gelectrondef}.

Due to the parity and the time reversal symmetry of the
Hamiltonian \eref{H} the functions \eref{Gholedef} and
\eref{Gelectrondef} obey the following symmetry properties
\begin{equation}
G_{h(e)}(x,t)=G_{h(e)}(-x,t)
\label{xtominusx}
\end{equation}
and
\begin{equation}
G_{h(e)}(x,t)=G_{h(e)}^*(x,-t).
\label{ttominust}
\end{equation}
Here the asterisk stands for the complex conjugation. Furthermore,
the following relation between $G_h$ and $G_e$ can be written
\begin{equation}
G_h(x,t)=-G_e^*(x,t)+ \left\langle
\left\{\psi_\uparrow^\dagger(x,t),\psi_\uparrow(0,0)\right\}\right
\rangle \label{GeGhanticom}
\end{equation}
where $\{,\}$ stands for the anticommutator. At $t=0$ this
relation becomes
\begin{equation}
G_h(x,0)=-G_e^*(x,0)+ \delta(x) \label{GeGhdelta}
\end{equation}
which is obvious from equation \eref{canonical}.
Note also the scaling relation
\begin{equation}
G_{e(h)}(x,t;\mu)=\sqrt{\mu} G_{e(h)}(\sqrt{\mu} x, \mu t;1)
\label{scaling}
\end{equation}
where $G_{e(h)}(x,t;\mu)$ is the correlation function taken at
a given chemical potential $\mu.$

In the following we will assume that $x\ge0,$ $t\ge 0$ and $\mu=1.$
The correlation functions for all other cases will follow
from the relations \eref{xtominusx}, \eref{ttominust} and \eref{scaling}.

\subsection{Determinant representation for $G_h$ and $G_e$ \label{detrep}}
In this subsection, following \cite{IP-98}, we write down the
determinant representation for the correlation functions
\eref{Gholedef} and \eref{Gelectrondef}. We use the notations
of \cite{GIKP-98}.

In paper \cite{IP-98} the following representation of the
correlation functions \eref{Gholedef} and \eref{Gelectrondef} was
derived
\begin{equation}
G_h(x,t)=\frac{\rme^{-\rmi t}}{8\pi} \int_{-\pi}^{\pi} \rmd \eta
F(\eta) B_{--}(x,t,\eta)\det(\hat I+\hat V)(x,t,\eta) \label{Gholeint}
\end{equation}
and
\begin{equation}
G_e(x,t)=-\frac{\rme^{\rmi t}}{2\pi} \int_{-\pi}^{\pi} \rmd \eta
\frac{F(\eta)}{1-\cos\eta} b_{++}(x,t,\eta)\det(\hat I+\hat V)(x,t,\eta).
\label{Gelectronint}
\end{equation}
To shorten notations we will henceforth omit $x,$ $t$ and $\eta$ in the arguments of
all functions unless they are necessary.

The objects entering the representations \eref{Gholeint} and \eref{Gelectronint} are
defined as follows. The function $F(\eta)$ is
\begin{equation}
F(\eta)=1+\frac{\rme^{\rmi\eta}}{2-\rme^{\rmi\eta}}
+\frac{\rme^{-\rmi\eta}}{2-\rme^{-\rmi\eta}}.
\end{equation}
The determinant
\begin{equation}
\fl\det (\hat I +\hat V) = \sum_{N=0}^{\infty}\frac1{N!}
\int_{-1}^1\rmd k_1\ldots\int_{-1}^1\rmd k_N\left|
\matrix{V(k_1,k_1)& \cdots & V(k_1,k_N)\cr \vdots & \ddots &
\vdots \cr V(k_N,k_1)& \cdots & V(k_N,k_N)} \right|
\end{equation}
is the Fredholm determinant of a linear integral operator~$\hat V$
with the kernel
\begin{equation}
V(k,p)=\frac{e_+(k)e_-(p)-e_+(p)e_-(k)}{k-p} \label{gammaVdef}
\end{equation}
defined on $[-1,1]\times[-1,1].$ The functions entering this
kernel are defined as follows:
\begin{equation}
\eqalign{ e_+(k)=\frac{1}{2\sqrt{\pi}}\rme^{-\tau(k)/2}
\left[(1-\cos\eta)\rme^{\tau(k)}E_0(k)+\sin\eta\right]\\
e_-(k)=\frac{1}{\sqrt{\pi}}\rme^{\tau(k)/2}\\
\tau(k)=\rmi k^2 t - \rmi k x\\
E_0(k)={\rm p.v.} \int_{-\infty}^\infty \rmd p
\frac{\rme^{-\tau(p)}}{\pi(p-k)}} \label{edefs}
\end{equation}
Note that $V(k,p)$ is symmetric, $V(k,p)=V(p,k),$ and nonsingular at $k=p.$

To define $B_{--}$ and $b_{++}$ introduce the resolvent operator
$\hat R:$
\begin{equation}
\hat I -\hat R=(\hat I +\hat V)^{-1}. \label{Resolvent}
\end{equation}
Due to the specific form \eref{gammaVdef} of the kernel $ V(k,p) $
the resolvent kernel $R(k,p)$ may be written as
\begin{equation}
R(k,p)=\frac{f_{+}(k)f_{-}(p)-f_{+}(p)f_{-}(k)}{k-p}
\label{gammaRdef}
\end{equation}
where functions $ f_{\pm}$ are the solutions to the integral
equations
\begin{equation}
f_{\pm}(k)+ \int_{-1}^{1} \rmd p V(k,p) f_{\pm}(p) = e_{\pm}(k).
\label{fpm}
\end{equation}
For the pairs $e_{\pm}(k) $ and $f_{\pm}(k)$ we will often use the
vector notation:
\begin{equation}
\vec e(k)=
\left(\begin{array}{c}
e_{+}(k) \\
e_{-}(k)
\end{array}\right) \qquad
\vec f(k)=
\left(\begin{array}{c}
f_{+}(k) \\
f_{-}(k)
\end{array}\right).
\label{vecfe}
\end{equation}
Now define the functions $B_{ab}$ and $C_{ab}$
\begin{equation}
B_{ab}=\int_{-1}^{1} \rmd k e_{a}(k) f_{b}(k) \qquad
C_{ab}=\int_{-1}^{1} \rmd k k e_{a}(k) f_{b}(k) \label{BabCab}
\end{equation}
where $a$ and $b$ run through two values: $ a, b = \pm$. In
particular,
\begin{equation}
B_{--}=\int_{-1}^1 \rmd k e_{-}(k)f_{-}(k). \label{Bmm}
\end{equation}
The function $b_{++}$ in \eref{Gelectronint} is
\begin{equation}
b_{++}=B_{++}-(1-\cos\eta)G_{0} \label{bpp}
\end{equation}
where $G_{0}(x,t)$ is the vacuum Green function
\begin{equation}
G_0(x, t)=\frac{1}{2 \pi} \int_{-\infty}^{\infty} \rmd k
\rme^{-\tau(k)}= \left\{\begin{array}{ll}
\frac{\rme^{-\rmi\pi/4}}{2\sqrt{\pi t}} \rme^{\rmi x^2/4 t}
&\qquad t\ne0
\\\delta(x) &\qquad t=0
 \end{array}\right. . \label{Gvac}
\end{equation}

\subsection{Deformation of the integration contour in equations \eref{Gholeint} and
\eref{Gelectronint} \label{residue}}
In this subsection we examine the analytic properties of the functions
entering the representations \eref{Gholeint} and
\eref{Gelectronint}. Using these properties we transform
the integrals in equations \eref{Gholeint} and
\eref{Gelectronint} to the form convenient for the asymptotical analysis.

Let us change the integration variable $\eta$ in \eref{Gholeint}
and \eref{Gelectronint} to
\begin{equation}
z=\rme^{\rmi \eta} \label{Zdef}.
\end{equation}
Then
\begin{equation}
G_h(x,t)=\frac{\rme^{-\rmi t}}{8\pi\rmi} \oint_{|z|=1} \frac{\rmd
z}{z} F(z) B_{--}(z)\det(\hat I+\hat V)(z) \label{Gholeintz}
\end{equation}
\begin{equation}
G_e(x,t)=\frac{\rme^{\rmi t}}{\pi\rmi} \oint_{|z|=1} \frac{\rmd
z}{(z-1)^2} F(z) b_{++}(z)\det(\hat I+\hat V)(z)
\label{Gelectronintz}
\end{equation}
where
\begin{equation}
\label{Fdef} F(z)=1+\frac{z}{2-z}+\frac{1}{2 z-1}.
\end{equation}
The integration contour, $|z|=1,$ is oriented counterclockwise.

Introduce the function
\begin{equation}
E(k) = \int_{-\infty}^{\infty}\frac{\rmd p}{2\pi}
\frac{\rme^{-\tau(p)}}{p-k} \qquad \mathrm{Im} k\ne 0 \label{Edef}
\end{equation}
and denote by $E_{+}(k)$ and $E_{-}(k)$ its analytic branches in
the upper and lower half-planes, respectively. The analytic
continuations of $E_{+}(k)$ and $E_{-}(k)$ to the real axis
satisfy
\begin{equation}
\fl E_{+}(k)+E_{-}(k)=E_0(k) \qquad E_{+}(k)-E_{-}(k)=\rmi
\rme^{-\tau(k)} \qquad k\in\mathbb R. \label{Epmprop}
\end{equation}
Using \eref{Epmprop} the kernel $V(k,p)=V(k,p;z)$ can be rewritten
as
\begin{eqnarray}
V(k,p;z)=(z-1)[W_{1}(k,p)+z^{-1} W_{2}(k,p)] \label{VEpm}
\end{eqnarray}
where
\begin{eqnarray}
W_{1}(k,p)=-\frac1{2\pi}\exp\left[\frac{\tau(k)+\tau(p)}2\right]
\frac{E_{+}(k)-E_{+}(p)}{k-p} \label{V1}\\
W_{2}(k,p)=\frac1{2\pi}\exp\left[\frac{\tau(k)+\tau(p)}2\right]
\frac{E_{-}(k)-E_{-}(p)}{k-p}. \label{V2}
\end{eqnarray}
By using the Fredholm theory of linear integral equations (see,
{\it e.g.}, \cite{CG-53}, chapter 3, section 7) the following two
facts may be shown: (a) the Fredholm determinant $\det(\hat I +
\hat V)(z)$ with the kernel $V(k,p;z)$ defined by
\eref{VEpm}-\eref{V2} is analytic in the complex $z$-plane except
at most at $z=0$ and $z=\infty.$ (b) the corresponding resolvent
kernel $R(k,p;z)$ may be represented as
\begin{equation}
R(k,p;z)=(z-1)\frac{D(k,p;z)}{\det(\hat I+\hat V)(z)} \label{R}
\end{equation}
where the function $D(k,p;z)$ is analytic in the complex $z$-plane
except at most at $z=0$ and $z=\infty.$ The explicit form of
$D(k,p;z)$ is given in the reference cited above. We will not use
it in our calculations.

\begin{figure}
\centering
\includegraphics[width=0.35\textwidth]{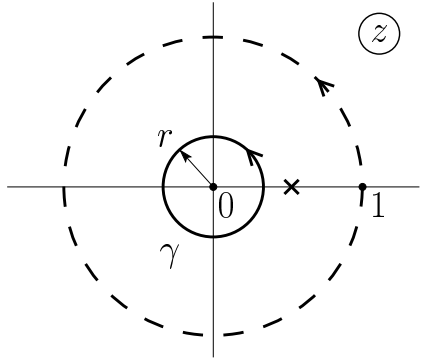}
\caption{Deformation of the integration contour. The dashed circle
shows the integration contour in equations \eref{Gholeintz} and
\eref{Gelectronintz}; the solid circle of radius $r<1/2$ shows the
integration contour $\gamma$ in equations \eref{Gholeintzdeformed}
and \eref{Gelectronintzdeformed}. Both contours are oriented
counterclockwise. The cross indicates the pole of the function
$F(z)$ at $z=1/2.$} \label{Fig:zintegral}
\end{figure}

Consider equation \eref{Gholeintz}. The function $B_{--},$
entering this equation, is defined by equation \eref{Bmm}.
Inverting \eref{fpm}
\begin{equation}
e_{-}(k)-\int_{-1}^1 \rmd p R(p,k;z) e_{-}(p) = f_{-}(k)
\end{equation}
and substituting the resulting expression in \eref{Bmm}, one gets
\begin{equation}
B_{--}(z)=\int_{-1}^{1}\rmd k e_{-}^2(k)-\int_{-1}^{1}\rmd
k\int_{-1}^{1}\rmd p e_{-}(k)R(k,p;z)e_{-}(p). \label{ber}
\end{equation}
Combining equations \eref{R} and \eref{ber} one sees that
$B_{--}(z)\det(\hat I +\hat V)(z)$ is analytic in the complex
$z$-plane except at most at $z=0$ and $z=\infty.$ Keeping this in
mind, deform the integration contour in \eref{Gholeintz} to a
contour $\gamma$ shown in figure~\ref{Fig:zintegral}.
Since the function $F(z)$ has a simple pole at the point $z=1/2$,
one gets
\begin{eqnarray}
\fl G_h(x,t)=\frac{\rme^{-\rmi t }}4 B_{--}(1/2)\det(\hat I+\hat
V)(1/2)\nonumber\\
\lo+ \frac{\rme^{-\rmi t }}{8\pi\rmi} \oint_{\gamma} \frac {\rmd
z} {z} F(z) B_{--}(z)\det(\hat I+\hat V)(z).
\label{Gholeintzdeformed}
\end{eqnarray}

Next consider equation \eref{Gelectronintz}. Like the function
$B_{--}(z)\det(\hat I+\hat V)(z),$ the function
$b_{++}(z)\det(\hat I +\hat V)(z)$ is analytic in the complex
$z$-plane except at most at $z=0$ and $z=\infty.$ Note that
despite the singular factor $(z-1)^{-2}$ the integrand does not
have a singularity at $z=1$ due to the second order zero of
$b_{++}(z)$ at this point. Indeed, one can see from \eref{edefs}
that $e_+\sim\eta$ for $\eta\sim0.$ Therefore, $f_{+}\sim\eta$ for
$\eta\sim0,$ which is seen from equation \eref{fpm}. Using the
definitions \eref{BabCab}, \eref{bpp} and \eref{Zdef} one can see
that $b_{++}\sim (1-z)^2$ for $1-z\sim 0$. Deforming the
integration contour in \eref{Gelectronintz} to the contour
$\gamma$ shown in figure~\ref{Fig:zintegral}, one gets
\begin{eqnarray}
\fl G_e(x,t)=4\rme^{\rmi t} b_{++}(1/2)\det(\hat I+\hat V)(1/2)\nonumber\\
\lo+ \frac{\rme^{\rmi t }}{\pi\rmi} \oint_{\gamma} \frac{\rmd
z}{(z-1)^2} F(z) b_{++}(z)\det(\hat I+\hat V)(z).
\label{Gelectronintzdeformed}
\end{eqnarray}

Consider a special case of $t=0.$ The kernel \eref{V2} becomes
equal to zero since $E_{-}(k)=0.$ Hence, the kernel \eref{VEpm}
becomes a linear function of $z.$ This implies that (see, {\it
e.g.}, \cite{CG-53}, chapter 3, section 7) (a) the determinant
$\det(\hat I+\hat V)(z)$ becomes an analytic function of $z$ for
all $z\in\mathbb C$ (b) the function $D(k,p;z)$ in equation
\eref{Resolvent} also becomes an analytic function of $z$ for all
$z\in\mathbb C.$ Thus, by virtue of \eref{ber} the function
$B_{--}(z)\det(\hat I+\hat V)(z)$ is analytic in the complex
$z$-plane and the second term in the right hand side of
\eref{Gholeintzdeformed} vanishes:
\begin{equation}
G_h(x,0)=\frac14 B_{--}(1/2)\det(\hat I+\hat V)(1/2).
\label{Gx0res}
\end{equation}
The function $G_e(x,0)$ can be obtained from equation
\eref{Gx0res} using equation \eref{GeGhdelta}.
\subsection{Differential equations \label{difur}}
In this subsection we write differential equations for the
Fredholm determinant $\det(\hat I+\hat V)(x,t,\eta)$ entering the
representations \eref{Gholeint} and \eref{Gelectronint}.

First, we obtain the differential equation with respect to $\eta.$
As follows from \eref{Zdef} and \eref{VEpm}-\eref{V2} the $\eta$
derivative of the kernel \eref{gammaVdef} is given by
\begin{equation}
\partial_{\eta}V(k,p)=
\frac{\rmi}{1-\rme^{-\rmi\eta}}V(k,p)+\frac{1-\rme^{-\rmi\eta}}{\rmi}W_2(k,p).
\label{diffkernel}
\end{equation}
Differentiating the identity
\begin{equation}
\ln\det (\hat I+\hat V) = \tr\ln (\hat I+\hat V)
\label{lndettrln}
\end{equation}
we immediately get
\begin{equation}
\fl\partial_\eta\ln\det(\hat I+\hat V)(x,t,\eta)=
\frac\rmi{1-\rme^{-\rmi\eta}}\tr[(\hat I+\hat V)^{-1}\hat V]+
\frac{1-\rme^{-\rmi\eta}}{\rmi} \tr[(\hat I+\hat V)^{-1}\hat W_2].
\label{diffetalndet}
\end{equation}
Transform the right hand side of \eref{diffetalndet} employing
\eref{Resolvent}, \eref{gammaRdef} and \eref{vecfe}. The result
for the first term is
\begin{equation}
\tr[(\hat I+\hat V)^{-1}\hat V]=\tr\hat R=
-\rmi \int_{-1}^{1} \rmd k [\vec f(k)]^{T}
\sigma_2 \partial _k \vec f(k) \label{detavect}
\end{equation}
where $\sigma_2$ is the second Pauli matrix
\begin{equation}
\fl \sigma_1=\left(\begin{array}{cc}
0&1 \\
1 & 0
\end{array}\right)
\qquad \sigma_2=\left(\begin{array}{cc}
0&-\rmi \\
\rmi & 0
\end{array}\right)
\qquad \sigma_3=\left(\begin{array}{cc}
1&0 \\
0 & -1
\end{array}\right).
\label{Pauli}
\end{equation}
The second term in the right hand side of \eref{diffetalndet} is
more complicated:
\begin{eqnarray}
\fl\tr[(\hat I+\hat V)^{-1}\hat W_2]&=\tr[(\hat I-\hat R) \hat W_2]\nonumber\\
&=\int_{-1}^{1}\rmd k W_2(k,k) - \int_{-1}^{1}\rmd k
\int_{-1}^{1}\rmd p R(k,p)W_2(p,k). \label{W}
\end{eqnarray}

The differential equations for $\det(\hat I + \hat V)(x,t,\eta)$
with respect to $x$ and $t$ can be obtained in a similar way. They
read
\begin{eqnarray}
\fl\partial_x \ln\det(\hat I +\hat V)(x,t,\eta) = \rmi B_{+-}=\rmi
B_{-+} \label{xderiv}
\\
\fl\partial_t\ln\det(\hat I+\hat V)(x,t,\eta)= - \rmi
\left[C_{+-}+C_{-+}+(1-\cos{\eta})G_0 B_{--}\right].
\label{tderiv}
\end{eqnarray}

Finally, the following initial conditions for the differential
equations follow from \eref{gammaVdef} and \eref{edefs}:
\begin{equation}
\det(\hat I +\hat V)(x,t,0)=1 \label{inicondeta}
\end{equation}
and
\begin{equation}
\det(\hat I +\hat V)(0,0,\eta)=1. \label{inicondxt}
\end{equation}

The differential equations \eref{diffetalndet}, \eref{xderiv} and
\eref{tderiv} will be used in the subsequent asymptotical
calculation of $\det(\hat I+\hat V).$
\subsection{Riemann-Hilbert problem \label{RHsection}}
In this subsection we briefly recall the notion of the
Riemann-Hilbert problem (RH problem) and its relation to the
objects entering the representations \eref{Gholeint} and
\eref{Gelectronint}.

\begin{figure} \centering
\includegraphics[width=0.5\textwidth]{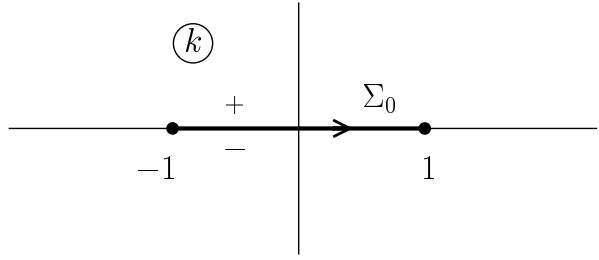}
\caption{The conjugation contour $\Sigma_0$. The thick line
connecting the points $k=-1$ and $k=1$ shows the conjugation
contour $\Sigma_0$ for the RH problem \eref{RHPtime}. The arrow
shows the direction of the orientation of this contour.}
\label{Fig:firstcontour}
\end{figure}

The standard formulation of the Riemann-Hilbert problem is as
follows \cite{DKMVZ-99,D-book,KBI-93}: for a given oriented
contour $\Sigma$ in the complex $k$-plane and for a given matrix
$\chi(k)$ defined on this contour, seek a matrix $Y(k)$ such that
\begin{eqnarray}
\label{RHPdef}
\eqalign{\mbox{(a)} \quad& Y(k) \mbox{ is analytic in } {\mathbb C}
\setminus\Sigma \\
\mbox{(b)} \quad& Y_+(k)=Y_-(k)\chi(k) \qquad  k \in\Sigma \\
\mbox{(c)} \quad& Y(k)\to I \mbox{ as } k\to\infty}
\end{eqnarray}
The contour $\Sigma$ is called `conjugation contour' or `jump
contour';  the matrix $\chi(k)$ is called `conjugation matrix' or
`jump matrix'. The matrices $Y_{\pm}(k)$ are the boundary values
of $Y(k)$ as $k$ approaches $\Sigma:$
\begin{equation}
Y_{\pm}(k)=\lim_{k'\to k}Y(k') \qquad \mbox{where} \qquad k'\in\pm
\mbox{ side of } \Sigma. \label{Ypm}
\end{equation}
By convention, the $+$ side (respectively, $-$ side) of $\Sigma$
lies to the left (respectively, right) as one traverses the
contour in the direction of the orientation. An example of such an
oriented contour is shown in figure~\ref{Fig:firstcontour}.

By using the Cauchy transform one can see that the RH problem
\eref{RHPdef} is equivalent to the matrix-valued integral equation
(\cite{KBI-93}, chapter XV)
\begin{equation}
Y(k)=I+\frac{1}{2\pi \rmi} \int_{\Sigma}\frac{\rmd s}{s-k}
Y_{-}(s)[\chi(s)-I]. \label{singeq}
\end{equation}
(The functions $Y_{\pm}(s),$ $\chi(s)$ and $Y_{\pm}(s)\chi(s)$ are
supposed to be integrable along the contour $\Sigma.$ We will
impose this condition for all RH problems appearing below).
Suppose the jump matrix $\chi(k)$ depends on the external
parameter, $x,$ and
\begin{equation}
\chi(k;x)=I+\mathcal O(x^{-\alpha}) \qquad \alpha\ge0 \qquad
\mbox{as} \qquad x\to\infty \label{conjas}
\end{equation}
uniformly for $k\in\Sigma$. Then, it can be shown solving
\eref{singeq} perturbatively that
\begin{equation}
Y=I+\mathcal O(x^{-\alpha}) \qquad \mbox{as} \qquad x\to\infty
\label{asympt}
\end{equation}
uniformly outside an arbitrarily small vicinity of the contour
$\Sigma.$ We will use the estimate \eref{asympt} in the subsequent
asymptotical analysis.

Next we relate the RH problem \eref{RHPdef} to the objects
entering the determinant representations \eref{Gholeint} and
\eref{Gelectronint}. Consider the following conjugation matrix
\begin{equation}
\nu_{0}(k;x,t)= \left(\begin{array}{cc} 1&0 \\
0&1\end{array}\right) -2\pi\rmi
\left(\begin{array}{cc}-e_-(k)e_+(k)&e_+^2(k)\\-e_-^2(k)&e_+(k)e_-(k)\end{array}\right)
\label{Gcondef}
\end{equation}
or in the vector notation \eref{vecfe}
\begin{equation}
\nu_0(k;x,t)=I+2\pi\vec e(k)[\vec e(k)]^{T}\sigma_2
\label{nu0vec}.
\end{equation}
Choose the conjugation contour to be the interval $[-1,1]$
oriented as shown in figure~\ref{Fig:firstcontour} and denote this
contour as $\Sigma_0.$
Let $Y$ be a solution to the RH problem
\begin{eqnarray}
\label{RHPtime}
\eqalign{\mbox{(a)} \quad& Y(k) \mbox{ is analytic in } \mathbb C\setminus\Sigma_{0} \\
\mbox{(b)} \quad& Y_+(k)=Y_-(k)\nu_{0}(k;x,t) \qquad k\in\Sigma_{0} \\
\mbox{(c)} \quad& Y(k)\to I \mbox{ as } k\to\infty}.
\end{eqnarray}
Then the components $f_{+}$ and $f_{-}$ of the vector
\begin{equation}
\vec f(k)=Y_{-}(k)\vec e(k)=Y_{+}(k)\vec e(k) \qquad k \in
\Sigma_0 \label{f=Ye}
\end{equation}
satisfy equations \eref{fpm}. Indeed, consider equation
\eref{singeq} where the conjugation matrix $\chi$ is chosen to be
the matrix $\nu_0$ given in equation \eref{nu0vec}. Multiply
equation \eref{singeq} by the vector $\vec e(k)$ from the right.
Equations \eref{fpm} follow immediately.

Combining equations \eref{f=Ye} and \eref{singeq} where the matrix
$\chi$ is chosen to be $\nu_0$ write
\begin{equation}
\fl Y(k)=\left(\begin{array}{cc} 1&0 \\ 0&1 \end{array}\right)+
\int_{\Sigma_0}\frac{\rmd p}{p-k}
\left(\begin{array}{cc}e_{-}(p)f_{+}(p) & -e_{+}(p)f_{+}(p) \\
e_{-}(p)f_{-}(p) & -e_{+}(p)f_{-}(p) \end{array}\right).
\label{RHPformalsolution}
\end{equation}
Comparing equations \eref{BabCab} and \eref{RHPformalsolution} one
sees that the functions $B_{ab}$ and $C_{ab}$ can be found from
the large $k$ expansion  of $Y(k)$
\begin{equation}
\fl Y(k)=\left(\begin{array}{cc} 1&0 \\ 0&1 \end{array}\right)+
\frac{1}{k}\left(\begin{array}{cc} -B_{-+} &B_{++} \\ -B_{--}
&B_{+-} \end{array}\right)+ \frac{1}{k^2}\left(\begin{array}{cc}
-C_{-+} &C_{++} \\ -C_{--} &C_{+-} \end{array}\right)+ {\cal
O}\left(\frac{1}{k^3}\right). \label{RHPexpansion}
\end{equation}

Consider an arbitrary compact domain $\mathcal D$ in the complex
plane of the parameter $z$ \eref{Zdef}, not containing the point
$z=0$. Since the Fredholm determinant in equations
\eref{Gholeintz} and \eref{Gelectronintz} is analytic in $\mathbb
C \setminus \{0\},$ it only has a finite set of zeroes
$\{z_n\}=\{z_1, \ldots, z_n\}$ in $\mathcal D.$ The integral
equations \eref{fpm} are solved uniquely for any $z\notin \{z_n\}$
in $\mathcal D.$ The solutions $f_{\pm}$ of the integral equations
\eref{fpm} can be found from a solution $Y(k)$ of the RH problem
\eref{RHPtime} by using equation \eref{f=Ye}. The functions
$B_{ab}$ and $C_{ab}$ can be found from the expansion
\eref{RHPexpansion}. The Fredholm determinant $\det(\hat I +\hat
V)(z)$ is completely defined by the differential equations
\eref{xderiv} and \eref{tderiv} with the initial condition
\eref{inicondxt} or, alternatively, by the differential equation
\eref{diffetalndet} with the initial condition \eref{inicondeta}.
Since the functions $B_{--}(z)\det(\hat I +\hat V)(z)$ in
\eref{Gholeintz} and $b_{++}(z)\det(\hat I +\hat V)(z)$ in
\eref{Gelectronintz} are analytic in $\mathbb C\setminus\{ 0\},$
their values at the points $z_1, \dots, z_n$ can be found by
analytic continuation.


\section{Riemann-Hilbert problem at $t=0$ \label{section:RHP0}}
In this section we solve the RH problem \eref{RHPtime}
asymptotically for $x\to\infty$ and $t=0.$ As it is obvious from
\eref{Gx0res}, to calculate $G_h(x,0)$ it is sufficient to solve
this RH problem at a specific value $z=1/2$ of the parameter $z.$
Nevertheless, we will construct the solution for all complex $z$
outside an arbitrarily small vicinity of the point $z=0$. This
general result will prove useful in the subsequent analysis of the
$t\ne0$ case.

In subsection \ref{RHP0} we formulate explicitly the RH problem
\eref{RHPtime} for $t=0,$ see equation \eref{RHPx}. In subsections
\ref{deformed1} and \ref{deformed2}, following along the lines of
\cite{DKMVZ-99,D-book}, we transform the RH problem \eref{RHPx} so
as to make it convenient for the large $x$ analysis. In subsection
\ref{explicit} we give an explicit solution to the jump relation
(\ref{RHPx}.b) in a vicinity of the end points $k=\pm 1$ of the
contour $\Sigma_0.$ In subsection \ref{perturthe} we construct the
solution to the RH problem \eref{RHPx} in the large $x$ limit by
matching the solution given in subsection \ref{explicit} with the
condition (\ref{RHPx}.c). In subsection \ref{approx} we summarize
the results of this section.

\subsection{Riemann-Hilbert problem \eref{RHPtime} at $t=0$ \label{RHP0}}

In this subsection we give an explicit formulation of the RH
problem \eref{RHPtime} at $t=0$.

By taking the limit $t\to 0$ and assuming that $x>0$ in
\eref{edefs}  we get for $e_{\pm}(k)$
\begin{equation}
e_{+}(k)=\frac{\rmi}2 \frac1{\sqrt\pi} \rme^{\rmi k
x/2}(1-\rme^{\rmi\eta})\\
e_{-}(k)=\frac1{\sqrt\pi}\rme^{-\rmi k x/2}. \label{epm0}
\end{equation}
The conjugation matrix \eref{Gcondef} becomes
\begin{equation}
\nu_0(k;x,0)=\left(\begin{array}{cc}
\rme^{\rmi \eta}& -2 \rmi \rme^{\rmi k x}\rme^{\rmi \eta}\sin^2\frac{\eta}2\\
2 \rmi \rme^{ - \rmi k x}&2-\rme^{\rmi \eta}
\end{array}\right) .
\label{nu0def}
\end{equation}
The oriented contour $\Sigma_0$ shown in
figure~\ref{Fig:firstcontour} remains unchanged. Thus, we have
defined the RH problem for $t=0$
\begin{eqnarray}
\label{RHPx}
\eqalign{\mbox{(a)}\quad& Y(k) \mbox{ is analytic in } \mathbb C\setminus\Sigma_{0} \\
\mbox{(b)}\quad& Y_+(k)=Y_{-}(k)\nu_0(k;x,0) \qquad k\in\Sigma_{0} \\
\mbox{(c)}\quad& Y(k)\to I \mbox{ as } k\to\infty.}
\end{eqnarray}

Introduce a matrix
\begin{equation}
P=\left(\begin{array}{cc}
0 &\rme^{\rmi \eta/2}\sin\frac{\eta}{2}
\\ \rme^{-\rmi \eta/2}\csc\frac{\eta}{2}&0
\end{array}\right).
\label{Pdef}
\end{equation}
Note that $P^2=I$. The jump matrix \eref{nu0def} and the functions
\eref{epm0} satisfy the involutions
\begin{equation}
P\nu_0(k)P^{-1}=[\nu_0(-k)]^{-1} \label{involution}
\end{equation}
and (note \eref{vecfe})
\begin{equation}
P\vec e(k)=\vec e(-k)
\end{equation}
respectively. As a consequence, we have the involution
\begin{equation}
PY(-k)P^{-1}=Y(k) \label{chiinvolution}
\end{equation}
for the solution $Y(k)$ of the RH problem \eref{RHPx} and, by virtue of \eref{f=Ye},
the involution
\begin{equation}
P \vec f(k)=\vec f(-k). \label{finvolution}
\end{equation}
for $\vec f(k).$

\subsection{Riemann-Hilbert problem on a lens-shaped contour \label{deformed1}}
\begin{figure}
\centering
\includegraphics[width=0.5\textwidth]{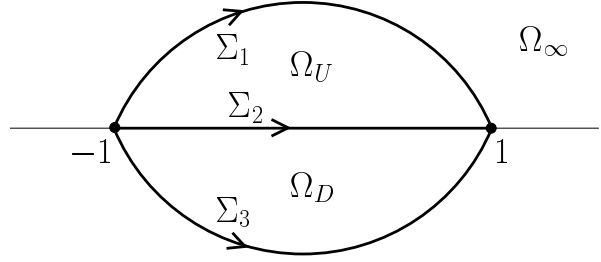}
\caption{The lens-shaped contour $\Sigma_\mathrm{lens}$. The thick
lines show the lens-shaped contour
$\Sigma_\mathrm{lens}$=$\Sigma_{1}\cup\Sigma_{2}\cup\Sigma_{3}.$
It is the conjugation contour for the RH problem
\eref{RHPxDeformed}. This contour divides the complex $k$-plane
into three regions: $\Omega_{U},$ $\Omega_{D}$ and
$\Omega_\infty.$ } \label{Fig:lens}
\end{figure}

In this subsection an RH problem equivalent to the RH problem
\eref{RHPx} is formulated on the lens-shaped conjugation contour
shown in figure \ref{Fig:lens}.

Represent $Y(k)$ as a product of two matrices
\begin{equation}
Y(k)=U(k)L(k) \label{Y=UL}
\end{equation}
and define $L(k)$ by the formula
\begin{equation}
L(k) =\left\{
\begin{array}{ll}
I &\qquad k\in\Omega_{\infty} \\
\nu_1(k) &\qquad k\in\Omega_{U}\\
\left[\nu_{3}(k)\right]^{-1} &\qquad k\in\Omega_{D}
\end{array}
\right.. \label{Ldef}
\end{equation}
A lens-shaped contour
$\Sigma_\mathrm{lens}=\Sigma_1\cup\Sigma_2\cup\Sigma_3$ dividing
the complex $k$-plane into the regions $\Omega_{U}$, $\Omega_{D}$
and $\Omega_{\infty}$ is displayed in figure~\ref{Fig:lens}.
The matrices $\nu_1$ and $\nu_3$ are defined as
\begin{eqnarray}
\nu_1(k;x)&=\left(\begin{array}{cc} 1& -2\rmi \rme^{\rmi kx}
\sin^2\frac{\eta}2\\ 0 &1
\end{array}\right) \vspace{20pt} \label{nu1def}
\\
\nu_3(k;x)&=\left(\begin{array}{cc} 1& 0\\ 2 \rmi \rme^{-\rmi k x}
\rme^{-\rmi \eta}&1 \end{array}\right). \label{nu3def}
\end{eqnarray}
Introducing a matrix $\nu_2$ by the formula
\begin{equation}
\nu_2=\left(\begin{array}{cc} \rme^{\rmi \eta}& 0\\
0&\rme^{-\rmi \eta} \end{array}\right) \label{nu2def}
\end{equation}
one can see that the following representation for the conjugation
matrix \eref{nu0def} can be written
\begin{equation}
\nu_0=\nu_3\nu_2\nu_1. \label{nusplit}
\end{equation}
Note that
\begin{equation}
P\nu_1(k)P^{-1}=[\nu_3(-k)]^{-1} \qquad P\nu_2P^{-1}=\nu_2^{-1}
\label{inv123}
\end{equation}
where the matrix $P$ is defined by \eref{Pdef}.

It follows from equations \eref{RHPx}, \eref{Y=UL}, \eref{Ldef}
and \eref{nusplit} that $U(k)$ solves the following RH problem
\begin{eqnarray}
\eqalign{\mbox{(a)}\quad& U(k) \mbox{ is  analytic  in } \mathbb C \setminus \Sigma_\mathrm{lens} \\
\mbox{(b)}\quad& U_{+}(k)=U_{-}(k)\nu_{i}(k) \qquad  k \in \Sigma_{i} \qquad i=1,2,3 \\
\mbox{(c)}\quad& U(k)\to I \mbox{ as } k\to \infty}
\label{RHPxDeformed}
\end{eqnarray}
We will calculate the large $x$ asymptotics of $U(k);$ the
asymptotics of $Y(k)$ will then follow from \eref{Y=UL}. The
advantage of the RH problem \eref{RHPxDeformed} from the point of
view of the large $x$ analysis will be clear from the next
subsection.

\subsection{Factorization of the Riemann-Hilbert problem
\eref{RHPxDeformed} \label{deformed2}}

In this subsection we present the scheme of the analysis of the RH
problem \eref{RHPxDeformed} in the large $x$ limit.

The $x$-dependent entry of the jump matrix $\nu_1$ (respectively,
$\nu_3$) entering \eref{RHPxDeformed} decays exponentially as
$x\to\infty$ everywhere on the jump contour $\Sigma_1$
(respectively, $\Sigma_3$) except at the points $k=\pm1.$ The jump
matrix $\nu_2$ is a constant matrix. For the RH problems with such
a behaviour of the jump matrices, the following scheme of the
large $x$ analysis can be employed \cite{DKMVZ-99,D-book}. Denote
by $\mathcal V$ a vicinity of the points $k=\pm 1.$ Split $U(k)$
in a product of three matrices
\begin{equation}
U=K S Q. \label{U=KSQ}
\end{equation}
The matrix $Q(k)$ satisfies the jump relation
$Q_{+}(k)=Q_{-}(k)\nu_2$ everywhere on the contour $\Sigma_2$ and
the jump relation (\ref{RHPxDeformed}.b) on the part of contour
$\Sigma_\mathrm{lens}$ lying in $\mathcal V.$ The matrix $S(k)$
represents the corrections to $Q(k)$ due to the mismatch between
the analytic branches of $Q(k)$ in and outside $\mathcal V.$ The
matrix $K(k)$ represents the corrections to $S(k)Q(k)$ due to the
discontinuities, exponentially small as $x\to\infty,$ across the
parts of the contours $\Sigma_1$ and $\Sigma_3$ lying
outside~$\mathcal V.$ Below we give the precise sense to $Q$, $S$
and $K.$

\begin{figure}
\centering
\includegraphics[width=0.5\textwidth]{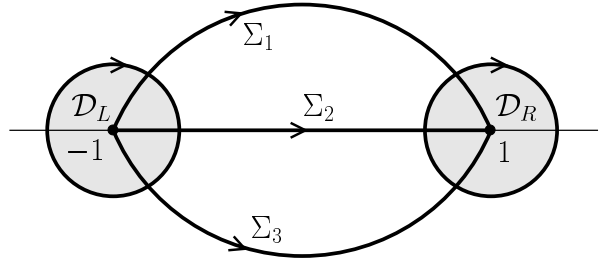}
\caption{The contour used in the factorization of the RH problem
\eref{RHPxDeformed}. The discs $\mathcal D_R$ and $\mathcal D_L$
are shaded in gray; their boundaries $\partial\mathcal D_R$ and
$\partial\mathcal D_L$ are oriented clockwise.}
\label{Fig:contourXLR}
\end{figure}

Let $\mathcal V$ consist of two discs $\mathcal D_R$ and $\mathcal
D_L$ shown in figure~\ref{Fig:contourXLR}. Let the radii of these
disks be equal and less than one. Denote the domain $\mathbb{C}
\setminus(\mathcal D_R\cup \mathcal D_L)$ as $\mathcal D_\infty.$
Introduce a matrix $Q(k)$ by
\begin{equation}
Q(k)=
\left\{
\begin{array}{cl} Q_R(k) &\qquad k \in \mathcal D_R \\
Q_L(k) &\qquad k \in \mathcal D_L \\
Q_{\infty}(k) &\qquad  k \in \mathcal D_\infty
\end{array}
\right. \label{Qdef}
\end{equation}
where the matrices $Q_\infty,$ $Q_R$ and $Q_L$ are defined as
follows. The matrix $Q_\infty(k)$ is analytic everywhere in
$\mathcal D_\infty$ except the contour $\Sigma_2 \cap \mathcal
D_\infty.$ On this contour it satisfies the jump relation with the
jump matrix \eref{nu2def}
\begin{equation}
\label{jumpofN} [Q_{\infty}(k)]_{+}=[Q_{\infty}(k)]_{-}\nu_2
\qquad  k\in \Sigma_2 \cap \mathcal D_\infty \label{Qinfjump}
\end{equation}
and is normalized at infinity:
\begin{equation}
Q_{\infty}(k)\to I\qquad \mbox{as}\qquad k\to \infty.
\label{Qinfnorm}
\end{equation}
We choose the following (obviously, not unique) solution to
\eref{Qinfjump} and \eref{Qinfnorm}:
\begin{equation}
Q_{\infty}(k)=
\left(\begin{array}{cc}
\left( \frac{k-1}{k+1}\right)^{\frac{\eta}{2 \pi}} & 0
\\ 0 &\left(\frac{k-1}{k+1}\right)^{-\frac{\eta}{2 \pi}}
\end{array}\right).
\label{Ndef}
\end{equation}
Note that it satisfies the involution
\begin{equation}
Q_\infty(k)=PQ_\infty(-k)P^{-1}. \label{Qinv}
\end{equation}
The matrix $Q_R(k)$ is analytic everywhere in $\mathcal D_R$
except the contour $\Sigma_\mathrm{lens}\cap\mathcal D_R.$ On this
contour it satisfies the jump relation with the jump matrices
\eref{nu1def}-\eref{nu2def}
\begin{equation}
\left[Q_R(k)\right]_{+}=\left[Q_R(k)\right]_{-} \nu_i(k) \qquad k
\in \Sigma_i\cap \mathcal D_R  \qquad i=1,2,3. \label{QRjump}
\end{equation}
A solution to the jump relation \eref{QRjump} is given in
subsection \ref{explicit}. The matrix $Q_L(k)$ is analytic
everywhere in $\mathcal D_L,$ except the contour
$\Sigma_\mathrm{lens}\cap\mathcal D_L.$ On this contour it
satisfies the jump relation similar to~\eref{QRjump}
\begin{equation}
\left[Q_L(k)\right]_{+}=\left[Q_L(k)\right]_{-} \nu_i(k) \qquad k
\in \Sigma_i\cap \mathcal D_L \qquad i=1,2,3. \label{QLjump}
\end{equation}
A solution to the jump relation \eref{QLjump} is given in
subsection \ref{explicit}.

It should be stressed that any solution to the jump relation
\eref{QRjump} (respectively, \eref{QLjump}) multiplied from the
left by an arbitrary matrix with the entries being entire
functions of $k$ in $\mathcal D_R$ (respectively, $\mathcal D_L$),
still remains a solution to \eref{QRjump} (respectively,
\eref{QLjump}). In subsequent calculations we will employ this
property to simplify the form of the conjugation matrices
$\theta_R$ and $\theta_L$ defined by equation \eref{thetaRLdef}
below.

Next define the matrix $S$ entering the decomposition \eref{U=KSQ}
as a solution to the following RH problem:
\begin{eqnarray}
\eqalign{\mbox{(a)}\quad& S(k) \mbox{ is analytic in }
\mathbb{C}\setminus(\partial \mathcal D_R\cup\partial \mathcal D_L) \\
\mbox{(b)}\quad& S_{+}(k)=S_{-}(k)\theta_{R,L} (k) \qquad  k \in \partial \mathcal D_{R,L}\\
\mbox{(c)}\quad& S(k)\to I \mbox{ as } k\to \infty} \label{RHPs}
\end{eqnarray}
with the clockwise oriented conjugation contours $ \partial
\mathcal D_{R,L}$ shown in figure~\ref{Fig:contourXLR}. The jump
matrices $\theta_{R,L}(k) $ are chosen to ensure the continuity of
$SQ$ across $\partial\mathcal D_{R,L}:$
\begin{equation}
S_{+}(k)Q_{+}(k)=S_{-}(k)Q_{-}(k) \qquad k \in \partial \mathcal D_{R,L}.
\end{equation}
This implies
\begin{equation}
\theta_{R,L}(k)=Q_{R,L}(k) [Q_{\infty}(k)]^{-1} \qquad
k\in\partial \mathcal D_{R,L}. \label{thetaRLdef}
\end{equation}
We solve the RH problem \eref{RHPs} in the large $x$ limit in
subsection \ref{perturthe}.

With the matrices $Q$ and $S$ defined above one can easily see
that the matrix $K$ in the decomposition \eref{U=KSQ} solves the
following RH problem
\begin{eqnarray}
\fl\eqalign{\mbox{(a)}\quad& K(k) \mbox{ is analytic in } \mathbb{C}\setminus(\Sigma_{1,3}\cap\mathcal D_\infty) \\
\mbox{(b)}\quad& K_{+}(k)=K_{-}(k)[S(k)Q(k)]\nu_{i}(k)[S(k)Q(k)]^{-1}
\qquad k \in\Sigma_i\cap\mathcal D_\infty \qquad i=1,3\\
\mbox{(c)}\quad& K(k)\to I \mbox{ as } k\to \infty} \label{K}
\end{eqnarray}
We solve the RH problem \eref{K} in the large $x$ limit in
subsection \ref{approx}.
\subsection{Exact solutions to the jump relations \eref{QRjump} and \eref{QLjump} \label{explicit}}
In this subsection we explicitly construct matrices $Q_R$ and
$Q_L$ satisfying the jump relations \eref{QRjump} and
\eref{QLjump}, respectively.

Since the conjugation matrix \eref{Gcondef} is $2\pi$-periodic in
$\mbox{Re}\eta$ we will restrict our considerations to the
interval
\begin{equation}
-\pi \le \mbox{Re}\eta \le \pi. \label{strip}
\end{equation}
To shorten notations, introduce the parameter
\begin{equation}
\epsilon=1-\frac{|\mbox{Re}\eta|}\pi.
\end{equation}

\begin{figure}
\centering
\includegraphics[width=0.3\textwidth]{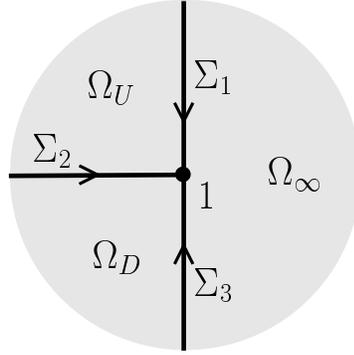}
\caption{The jump contour for the jump relation \eref{QRjump}. The
disk $\mathcal D_R$ is shaded in gray. The segments of the
contours $\Sigma_1$ and $\Sigma_3$ lying in $\mathcal D_R$ are
straight and perpendicular to $\Sigma_2.$ } \label{Fig:contourR}
\end{figure}

Consider the part of the contour $\Sigma_\mathrm{lens}$ lying in
the disk $\mathcal D_R.$ For the convenience of presentation we
choose $\Sigma_1$ and $\Sigma_3$ to be perpendicular to
$\Sigma_2,$ as shown in figure~\ref{Fig:contourR}.
Consider a matrix $\Phi_R(k;x)$ with the entries
\begin{equation}
\fl\eqalign{\Phi^R_{11}(k;x)&= x^{-\frac{\eta}{2 \pi}}
\varsigma_1(k-1)\rme^{\frac{\rmi \eta}{4}}
\Psi\left[-\frac{\eta}{2 \pi},1;-\rmi x(k-1)\right]
\\
\Phi^R_{21}(k;x) &=- x^{\frac{\eta}{2\pi}} \frac
{\pi\varsigma_1(k-1) \rme^{-\frac{\rmi \eta}{4}}\rme^{-\rmi x}} {
\Gamma^2\left(\frac{\eta}{2\pi}\right) \sin^2\frac{\eta}{2}}
\Psi\left[1-\frac{\eta}{2\pi}, 1; -\rmi x(k-1)\right]
\\
\Phi^R_{12}(k;x)&= x^{-\frac{\eta}{2 \pi}} \frac {\pi
\varsigma_2(k-1)\rme^{\frac{3 \rmi \eta}{4}}\rme^{\rmi x}}
{\Gamma^2\left(-\frac{\eta}{2 \pi}\right)}
\Psi\left[1+\frac{\eta}{2\pi},1;\rmi x(k-1)\right]
\\
\Phi^R_{22}(k;x)&= x^{\frac{\eta}{2 \pi}}
\varsigma_2(k-1)\rme^{\frac{\rmi \eta}{4}}
\Psi\left[\frac{\eta}{2\pi},1;\rmi x(k-1)\right] }
\label{phientries}
\end{equation}
Here $\Psi(a, b; w)$ is the Tricomi $\Psi$ function, which has a
branch cut discontinuity across the negative real axis
\cite{BE-53}
\begin{eqnarray}
\fl\Psi(a, 1, w+\rmi 0)-\left[1-2 \rmi \rme^{-\rmi a \pi} \sin(\pi a) \right]\Psi(a,1, w- \rmi 0) \nonumber \\
\lo= \frac{2\pi \rmi \rme^{- \rmi a \pi}}{\Gamma^2(a)}
e^w\Psi(1-a, 1, -w) \qquad w\in \mathbb{R}_{-} \label{Tricomi}
\end{eqnarray}
and $\varsigma_{1,2}(w)$ are piecewise-constant functions defined
by
\begin{equation}
\varsigma_1(w)=\left\{
\begin{array}{ll}
\rme^{-\rmi \eta} & \qquad -\pi<\arg w<-\frac{\pi}{2}\\
1 & \qquad -\frac{\pi}{2}<\arg w<\pi
\end{array}
\right. \label{varsigma1}
\end{equation}
and
\begin{equation}
\varsigma_2(w)=\left\{
\begin{array}{ll}
1 & \qquad -\pi <\arg w< \frac{\pi}{2} \\
\rme^{-\rmi \eta} & \qquad  \frac{\pi}{2}<\arg w< \pi
\end{array}
\right. . \label{varsigma2}
\end{equation}
Using \eref{Tricomi} it is easy to check that $\Phi_R(k;x)$
satisfies the jump relation \eref{QRjump} on the contour
$\Sigma_\mathrm{lens}\cap\mathcal D_R$ shown in
figure~\ref{Fig:contourR}
\begin{equation}
\fl[\Phi_R(k;x)]_{+}=[\Phi_R(k;x)]_{-}\nu_i(k;x) \qquad
k\in\Sigma_{i}\cup\mathcal D_R \qquad i=1,2,3. \label{PhiRjump}
\end{equation}
It follows from the involution \eref{inv123} that a solution
$\Phi_L(k)$ to the jump relation \eref{QLjump} may be obtained
from $\Phi_R(k)$ via the involution
\begin{equation}
\Phi_L(k;x)=P\Phi_R(-k;x)P^{-1} \qquad k \in \mathcal D_L. \label{PhiRtoL}
\end{equation}

As was mentioned in the paragraph following equation
\eref{QLjump}, any matrix $Q_R$ of the form
\begin{equation}
Q_R(k)=E(k)\Phi_R(k) \label{AnsaR}
\end{equation}
with $E(k)$ being an arbitrary analytic in $\mathcal D_R$ matrix,
satisfies \eref{QRjump}. Choosing
\begin{equation}
E(k)=\left(
\begin{array}{cc}
(k+1)^{ -\frac{\eta}{2 \pi} } &0 \\
0 & (k+1)^{\frac{\eta}{2\pi}}
\end{array}
\right) \label{E}
\end{equation}
one ensures that $Q_R(k)$ goes to $Q_\infty(k),$ as $x\to\infty,$
for any $\epsilon\ne0,$ see \eref{PhiRasympt0}. This will
facilitate the asymptotic solution of the RH problem \eref{RHPs}.
Taking into account \eref{PhiRtoL} and \eref{AnsaR} define
$Q_L(k)$ in $\mathcal D_L$ by the formula
\begin{equation}
Q_L(k)=PQ_R(-k)P^{-1} \qquad k\in\mathcal D_L. \label{RtoL}
\end{equation}

To calculate the large $x$ asymptotics of $Q_R(k)$ we use the
asymptotic formula for the Tricomi $\Psi$ function (\cite{BE-53},
section 6.13.1)
\begin{equation}
\Psi(a, 1, w)=w^{-a}\left[1-\frac{a^2}{w}+ {\cal O}\left(\frac 1
{w^2}\right)\right] \qquad \mbox{as} \qquad w\to \infty.
\label{Tricomias}
\end{equation}
Applying \eref{Tricomias} to \eref{phientries} one gets from
\eref{AnsaR}:
\begin{equation}
\fl Q_R(k)=Q_{\infty}(k)+\frac{1}{x(k-1)}
\left(\begin{array}{cc}
-\rmi \left(\frac{\eta}{2\pi}\right)^2  \left(
\frac{k-1}{k+1}
\right)^{\frac {\eta} {2 \pi}}
&
a_{R}x^{-\frac{\eta}{\pi}} (k^2-1)^{-\frac{\eta}{2\pi}}
 \\
b_{R}x^{\frac{\eta}{\pi}}(k^2-1)^{\frac{\eta}{2\pi}}
&
\rmi \left(\frac{\eta}{2\pi}\right)^2
\left(
\frac{k+1}{k-1}
\right)^{\frac {\eta} {2 \pi}}
\end{array}\right)
+{\cal O} \left( x^{-1-\epsilon} \right) \label{PhiRasympt0}
\end{equation}
where
\begin{equation}
a_R= -\frac{\rmi \pi \rme^{\rmi \frac\eta2}\rme^{\rmi x}}{\Gamma^2\left(-\frac{\eta}{2 \pi}\right)}
\qquad
b_R=-\frac{\rmi \pi \rme^{-\rmi \frac \eta 2}\rme^{-\rmi x }}
{\Gamma^2\left(\frac\eta{2\pi}\right)\sin^2\frac\eta2}.
\end{equation}
The large $x$ asymptotics of $Q_{L}(k)$ follows from \eref{RtoL}
and \eref{PhiRasympt0}. Note that the terms in asymptotic
expansions of $Q_{R}$ explicitly written down in
\eref{PhiRasympt0} do not have jumps across the contours
$\Sigma_1$ and $\Sigma_3$. This reflects the fact that the
$x$-dependent entries in the jump matrices $\nu_1$ and $\nu_3$ are
exponentially small on $\Sigma_1$ and $\Sigma_3,$ respectively.
\subsection{Solution to the Riemann-Hilbert problem \eref{RHPs} in the large $x$ limit \label{perturthe} }

In this subsection we solve the RH problem \eref{RHPs} in the
large $x$ limit.

Consider the conjugation matrices $\theta_R$ and $\theta_L$ of the
RH problem \eref{RHPs}. They are defined by the formula
\eref{thetaRLdef} and satisfy the involution
\begin{equation}
\theta_L(k)=P\theta_R(-k)P^{-1} \qquad k \in\partial \mathcal D_L \label{thetainv}
\end{equation}
as can be shown using the involutions \eref{Qinv} and
\eref{RtoL}. Substituting the asymptotic expression
\eref{PhiRasympt0} into \eref{thetaRLdef} one gets the
approximation of the matrix $\theta_R$ by a matrix
$\tilde\theta_R$
\begin{equation}
\theta_R=\tilde\theta_R[I+\mathcal O(x^{-1-\epsilon})] \label{thetaR}
\end{equation}
uniform for $k\in\partial\mathcal D_R.$ Here
\begin{equation}
\tilde\theta_R(k)=I+\frac{\upsilon_R(k)}{x(k-1)} \qquad k\in
\partial \mathcal D_{R} \label{thetaRas}
\end{equation}
and
\begin{equation}
\upsilon_R(k)= \left(\begin{array}{cc}
-\rmi\left(\frac{\eta}{2\pi}\right)^2 &
a_R[x(k+1)]^{-\frac\eta\pi}
\\
b_R [x(k+1)]^{\frac\eta\pi} &
\rmi\left(\frac{\eta}{2\pi}\right)^2
\end{array}\right).
\label{upsilonR}
\end{equation}
It follows from the involution \eref{thetainv} that the matrix
$\theta_L$ is approximated by a matrix $\tilde\theta_L$
\begin{equation}
\theta_L=\tilde\theta_L[I+\mathcal O(x^{-1-\epsilon})] \label{thetaL}
\end{equation}
uniformly for $k\in\partial\mathcal D_L,$ where
\begin{equation}
\tilde\theta_L(k)=P\tilde\theta_R(-k)P^{-1} \qquad k\in
\partial \mathcal D_{L}. \label{thetaLas}
\end{equation}

Consider the RH problem
\begin{eqnarray}
\label{RHPsl} \eqalign{\mbox{(a)}\quad& \tilde S(k) \mbox{ is
analytic in }
\mathbb{C}\setminus(\partial \mathcal D_R\cup\partial \mathcal D_L) \\
\mbox{(b)}\quad& \tilde S_{+}(k)=\tilde S_{-}(k)\tilde\theta_{R,L} (k) \qquad k \in \partial \mathcal D_{R,L}\\
\mbox{(c)}\quad& \tilde S(k)\to I \mbox{ as } k\to \infty}
\end{eqnarray}
with the  matrices $\tilde\theta_{R}$ and $\tilde\theta_{L}$
defined by equations \eref{thetaRas} and \eref{thetaLas},
respectively. Comparing \eref{thetaR} and \eref{thetaL} with
\eref{conjas} and \eref{asympt} we conclude that the solution
$\tilde S(k)$ of the RH problem \eref{RHPsl} approximates the
solution $S(k)$ of the RH problem \eref{RHPs}
\begin{equation}
S=\tilde S [I+\mathcal O(x^{-1-\epsilon})] \label{Stilde}
\end{equation}
uniformly outside a vicinity of the conjugation contours
$\partial\mathcal D_R$ and $\partial\mathcal D_L.$

Due to a simple analytical structure of the conjugation matrices
\eref{thetaRas} and \eref{thetaLas} the RH problem \eref{RHPsl}
can be solved exactly. To begin with, introduce the following
notation for the analytic branches of $\tilde S(k)$
\begin{equation}
\tilde S(k)=
\left\{
\begin{array}{cl}
\tilde S_R(k) &\qquad k \in \mathcal D_R \\
\tilde S_L(k) &\qquad k \in \mathcal D_L \\
\tilde S_{\infty}(k) &\qquad k \in \mathcal D_\infty
\end{array}\right..
\label{sdef}
\end{equation}
The matrix $\tilde\theta_R(k),$ analytically continued from
$\partial \mathcal D_R$ into $\mathcal D_R$ has a simple pole at
$k=1,$ as follows from the representation \eref{thetaRas}. The
matrix $\tilde S_{R}(k)$ is analytic in $\mathcal D_R.$ Therefore,
the jump relation (\ref{RHPsl}.b)
\begin{equation}
\tilde S_{\infty}(k)=\tilde S_{R}(k) \tilde\theta_{R}(k) \qquad
k\in\partial \mathcal D_R \label{jrR}
\end{equation}
implies that the analytic continuation of $\tilde S_{\infty}(k)$
into $\mathcal D_R$ has a simple pole at $k=1.$ Similarly, the
analytic continuation of $\tilde S_{\infty}(k)$ into $\mathcal
D_L$ has a simple pole at $k=-1.$ Finally, taking into account the
condition given in equation (\ref{RHPsl}.c), $\tilde
S_{\infty}(\infty)=I$, we see that the most general form of
$\tilde S_{\infty}(k)$ is
\begin{equation}
\tilde S_{\infty}(k)=I+ \frac{A_R}{k-1}+\frac{A_L}{k+1} \qquad
k\in\mathbb{C} \label{scgeneral}
\end{equation}
with the matrices $A_{R}$ and $A_L$ being independent of $k$.

The involution
\begin{equation}
P\tilde S_{\infty}(k)P^{-1}=\tilde S_{\infty}(-k)
\end{equation}
which follows from the involution \eref{thetaLas}, provides a
simple relation between $A_{R}$ and $A_{L}$
\begin{equation}
A_{R}=-PA_{L}P^{-1} \label{ARtoAL}
\end{equation}
so we only need to calculate one of the matrices, say, $A_{R}$.
For this, invert \eref{jrR}:
\begin{equation}
\tilde S_{R}(k)=\tilde S_{\infty}(k) [\tilde\theta_{R}(k)]^{-1}
\qquad k\in \partial \mathcal D_R. \label{jrRinverse}
\end{equation}
Using \eref{thetaRas} and \eref{scgeneral} the right hand side of
equation \eref{jrRinverse} can be analytically continued into
$\mathcal D_R.$ In the following, equation \eref{jrRinverse} will
be understood to be such a continuation. The matrix
$[\tilde\theta_{R}(k)]^{-1}$ has the form
\begin{equation}
[\tilde\theta_{R}(k)]^{-1} = [\det\tilde\theta_{R}(k)]^{-1}
\left[I-\frac{\upsilon_{R}(k)}{x(k-1)}\right]
\end{equation}
where
\begin{equation}
\det\tilde\theta_{R}(k) =1+\frac{\det\upsilon_{R}}{x^2(k-1)^2}
\end{equation}
and $\det\upsilon_{R}$ is a $k$-independent number
\begin{equation}
\det\upsilon_{R} = \left(\frac{\eta}{2\pi}\right)^{4} + \left(\frac{\eta}{2\pi}\right)^{2}.
\end{equation}
The right hand side of \eref{jrRinverse} must be analytic in
$\mathcal D_R,$ since the left hand side is. The matrix
$[\tilde\theta_{R}(k)]^{-1}$ goes to zero as $k\to1,$ cancelling
the pole of $\tilde S_{\infty}(k)$ at this point. However,
$[\tilde\theta_R(k)]^{-1}$ is singular at the points where
\begin{equation}
\det\tilde\theta_{R}(k) = 0. \label{quadratic}
\end{equation}
Equation \eref{quadratic} is quadratic in $k$. Its solutions,
$k^{R}_{1,2},$ lie inside the disk $\mathcal D_R$ for sufficiently
large $x$. Therefore, to cancel the singularities in
\eref{jrRinverse} the matrix $A_{R}$ has to satisfy the pair of
matrix equations
\begin{equation}
\left(I + \frac{A_R}{k_{1,2}^{R}-1}
-\frac{P^{-1}A_{R}P}{k_{1,2}^{R}+1} \right) \left[I -
\frac{\upsilon_{R}(k_{1,2}^R)}{x(k_{1,2}^R-1)} \right]=0.
\label{eqA}
\end{equation}
The determinant of the matrix in square brackets  in \eref{eqA} is
equal to $\det\tilde\theta_{R}(k^{R}_{1,2})$ and, therefore, is
equal to zero. Employing this fact one gets (after some algebra)
the matrix~$A_R.$ The large $x$ asymptotics of $A_R$ is
\begin{equation}
A_R=\frac{2}\kappa \left(\begin{array}{cc}
\varkappa_L^2-\frac{\rmi}{2 x} \left(\frac {\eta}{2 \pi} \right)^2
& -\rmi \varkappa_R \rme^{\rmi \frac{\eta}{2}} \sin \frac{\eta}{2}
\\
-\rmi\varkappa_L \rme^{-\rmi\frac\eta2} \csc\frac{\eta}{2} &
\varkappa_R^2+\frac{\rmi}{2 x} \left(\frac{\eta}{2\pi} \right)^2
\end{array}\right)
+ \mathcal O(x^{-1-\epsilon}) \label{answerAR}
\end{equation}
where
\begin{equation}
\varkappa_R=\frac{\pi \rme^{\rmi x}
(2x)^{-1-\frac\eta\pi}}{\Gamma^2\left(-\frac\eta{2\pi}\right)
\sin\frac \eta 2 }\qquad \varkappa_L=\frac{\pi\rme^{-\rmi x}
(2x)^{-1+\frac\eta\pi}}{\Gamma^2\left(\frac\eta{2\pi}\right)
\sin\frac \eta 2} \label{kappadef}
\end{equation}
and
\begin{equation}
\kappa=1+\varkappa_R^2+\varkappa_L^2. \label{Delta}
\end{equation}
Applying the involution \eref{ARtoAL} to \eref{answerAR} one gets
the large $x$ asymptotics of $A_L:$
\begin{equation}
A_L=\frac{2}\kappa \left(\begin{array}{cc}
-\varkappa_R^2-\frac{\rmi}{2 x} \left(\frac {\eta}{2 \pi}
\right)^2 & \rmi \varkappa_L \rme^{\rmi \frac{\eta}{2}} \sin
\frac{\eta}{2}
\\
\rmi\varkappa_R \rme^{-\rmi \frac{\eta}{2}} \csc\frac{\eta}{2}
 &
-\varkappa_L^2+\frac{\rmi}{2 x} \left(\frac {\eta}{2 \pi}
\right)^2
\end{array}\right)+ \mathcal O(x^{-1-\epsilon}).
\label{answerAL}
\end{equation}

\subsection{Riemann-Hilbert problem \eref{RHPx}: results in the large $x$ limit
\label{approx}} In this subsection we show that the the matrix $K$
in the decomposition \eref{U=KSQ} does not contribute to the
asymptotic solution of the RH problem \eref{RHPxDeformed} to the
order we are interested in. This will conclude the asymptotic
solution of the RH problem \eref{RHPx}.

Consider the decomposition \eref{U=KSQ}. In subsection
\ref{perturthe} we constructed explicitly the approximation
$\tilde S$ to $S$ up to the order of $x^{-1-\epsilon}.$ This
approximation is uniform outside an arbitrarily small vicinity of
the conjugation contours $\partial\mathcal D_R$ and
$\partial\mathcal D_L.$ The matrix $Q$ was constructed explicitly
in subsections \ref{deformed2} and \ref{explicit}. Using these
results, solve asymptotically the RH problem \eref{K} for the
matrix $K.$ The matrices $\nu_1$ and $\nu_3,$ entering
(\ref{K}.b), converge as $x\to\infty$ to the identity matrix
uniformly and exponentially fast on the contours
$\Sigma_1\cap\mathcal D_\infty$ and $\Sigma_3\cap\mathcal
D_\infty$, respectively. One can prove that
\begin{equation}
K=I+o(x^{-\infty}) \label{Kas}
\end{equation}
uniformly outside a vicinity of the contours
$\Sigma_{1,3}\cap\mathcal D_\infty$. The basic idea in obtaining
\eref{Kas} is the same as in obtaining \eref{Stilde}: to use
\eref{singeq}. One should take into account, though, that we have
now, contrary to \eref{Stilde}, no uniform estimate for the jump
matrix in (\ref{K}.b) in a vicinity of the contours
$\partial\mathcal D_{R,L}$ since we have no uniform estimate for
$S(k)$ entering this matrix, in this vicinity. For further details
we refer the reader to reference \cite{DKMVZ-99}.

Substituting \eref{Stilde} and \eref{Kas} into the decomposition
\eref{U=KSQ} one gets the estimate
\begin{equation}
U=\tilde SQ[I+\mathcal O(x^{-1-\epsilon})] \label{schi}
\end{equation}
uniform outside an arbitrarily small vicinity of the contours
$\partial\mathcal D_{R,L}$ and $\Sigma_{1,3}\cap\mathcal
D_\infty.$ Using \eref{Y=UL} one gets the approximation to $Y(k)$
up to the order of $x^{-1-\epsilon}$, uniform outside this
vicinity. This completes the solution to the RH problem
\eref{RHPx} in the large $x$ limit.

\section{Equal-time correlation functions \label{zerotimeG}}
In this section we calculate the equal-time correlation functions
$G_h(x,0)$ and $G_e(x,0)$ in the large $x$ limit.  The asymptotic
expressions for these functions were obtained in papers
\cite{BL-87,B-91}. The main purpose of this section is a to give a
detailed illustration of our method on this known example. The
procedures of this section will be extensively used in the
subsequent analysis of the time-dependent correlation functions.

In subsection \ref{Basympt} we calculate the functions $B_{ab}$
\eref{BabCab} from the large $k$ expansion \eref{RHPexpansion} of
$Y(k).$ From thus obtained $B_{+-}$ we calculate the Fredholm
determinant $\det(\hat I+\hat V)$ in the large $x$ limit, see
equation \eref{logdet}. The constant $C(\eta)$ entering
\eref{logdet} is calculated in subsection \ref{casy}. We give the
answer for $G_{h}(x,0)$ and $G_{e}(x,0)$ in subsection
\ref{answer0}.

\subsection{Large $x$ asymptotics of $B_{ab}$ and of the Fredholm determinant $\det(\hat I+\hat
V)$\label{Basympt}}

In this subsection we use the results of section
\ref{section:RHP0} on the  asymptotic solution of the RH problem
\eref{RHPx} to calculate the functions $B_{ab}$ and the Fredholm
determinant $\det(\hat I+\hat V)$ in the large $x$ limit.

It follows from \eref{schi}, \eref{Y=UL}, \eref{Qdef} and
\eref{sdef} that in the vicinity of $k=\infty$ the following
uniform approximation is valid:
\begin{equation}
Y=\tilde S_{\infty}Q_{\infty}+{\cal O}(x^{-1-\epsilon}).
\label{SQinfty}
\end{equation}
Recall that the matrix $Q_\infty(k)$ is given by the formula
\eref{Ndef}, the matrix $\tilde S_\infty(k)$ by the formulas
\eref{scgeneral}, \eref{answerAR} and \eref{answerAL}. The
approximation \eref{SQinfty} being uniform, the functions $B_{ab}$
can be calculated from the large $k$ expansion \eref{RHPexpansion}
of equation \eref{SQinfty}. Expanding \eref{Ndef} and
\eref{scgeneral} in inverse powers of $k$ and comparing with
\eref{RHPexpansion} one gets
\begin{equation}
\left(\begin{array}{cc} -B_{-+} &B_{++} \\
-B_{--} &B_{+-}\end{array}\right)= A_R+A_L -\frac{\eta}{\pi}
\sigma_3 +\mathcal O \left(x^{-1-\epsilon}\right)
\label{Yexpansion}
\end{equation}
where $\sigma_3$ is the third Pauli matrix \eref{Pauli}.
Substituting equations \eref{answerAR} and \eref{answerAL} into
\eref{Yexpansion} one gets for $B_{+-}$ and $B_{--}$
\begin{eqnarray}
B_{+-}&=\frac{\eta}{\pi}+ \frac2\kappa
\left[\varkappa_R^2-\varkappa_L^2+\frac\rmi{x}
\left(\frac\eta{2\pi}\right)^2\right] +\mathcal O
\left(x^{-1-\epsilon}\right) \label{Bpmas}
 \\
B_{--}&=2 \rmi\rme^{-\frac{\rmi\eta}2} \csc\frac{\eta}2
(\varkappa_L- \varkappa_R)[1 +\mathcal O
\left(x^{-\epsilon}\right)] \label{Bmmas}
\end{eqnarray}
where $\varkappa_{R}$ and $\varkappa_{L}$ are defined by
\eref{kappadef} and $\kappa$ is defined by \eref{Delta}.

To calculate the Fredholm determinant $\det(\hat I+\hat V)$ we
integrate \eref{xderiv} taking into account the boundary condition
\eref{inicondxt}:
\begin{equation}
\ln \det(\hat I + \hat V)=  \rmi \int_0^x \rmd y B_{+-}(y).
\label{xintegrate}
\end{equation}
Substituting equation \eref{Bpmas} into \eref{xintegrate} and
calculating the integral in the large $x$ limit, one obtains
\begin{equation}
\ln{\rm det}(\hat I + \hat V)= \frac{\rmi \eta}{\pi} x
-2\left(\frac{\eta}{2\pi}\right)^2 \ln x+ C(\eta)+ {\cal
O}(x^{-\epsilon}). \label{logdet}
\end{equation}
We calculate the integration constant $C(\eta)$ in subsection
\ref{casy}.

\subsection{Calculation of $C(\eta)$ \label{casy}}

In this subsection we calculate the constant $C(\eta)$ in
\eref{logdet} using the differential equation \eref{diffetalndet}.

Consider equation \eref{diffetalndet}. For $t=0$ the kernel
$W_2(k,p)$ vanishes and, therefore, the second term in the right
hand side of \eref{diffetalndet} vanishes. The remaining term is
given by equation \eref{detavect}. With the help of the involution
\eref{finvolution} and the identity
\begin{equation}
P^T\sigma_2 P = - \sigma_2 \label{sigma2involution}
\end{equation}
the integral in \eref{detavect} can be reduced to the integral
over positive~$k,$ yielding
\begin{equation}
\partial_\eta\ln\det(\hat I +\hat V)(x,0,\eta)=
\frac{2}{1-\rme^{-\rmi \eta}} \int_{0}^{1} \rmd k [\vec f (k)]^{T}
\sigma_2 \partial _k \vec f(k). \label{detaposik}
\end{equation}

\begin{figure}
\centering
\includegraphics[width=0.5\textwidth]{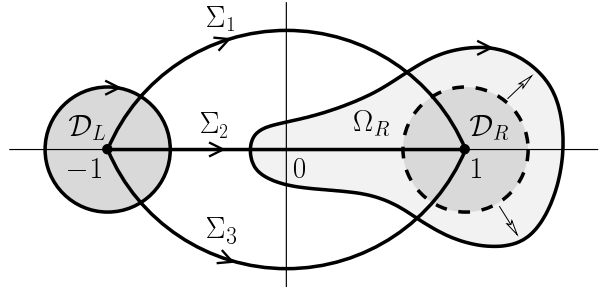}
\caption{Analytic continuation. Equation \eref{schi} is
analytically continued from the disk $\mathcal D_R$ into the
region $\Omega_R$ containing the interval $[0,1].$}
\label{Fig:OmegaRcontinuation}
\end{figure}

The large $x$ asymptotics of the function $\vec f(k)$ in equation
\eref{detaposik} is calculated as follows. Use equation
\eref{f=Ye}
\begin{equation}
\vec f(k)=Y_{+}(k)\vec e(k) \qquad k\in\Sigma_2 \label{f=ye}
\end{equation}
and replace $Y(k)$ by its large $x$ asymptotics obtained in
section \ref{section:RHP0}. It follows from \eref{Y=UL} that
\begin{equation}
Y_{+}(k)=U_{+}(k) \nu_{1}(k) \qquad k\in\Sigma_2 \label{Y=Unu}
\end{equation}
where $U(k)$ solves the RH problem \eref{RHPxDeformed}. The
formula \eref{schi} gives the approximation to $U(k)$ uniform
outside an arbitrarily small vicinity of the contours
$\partial\mathcal D_{R,L}$ and $\Sigma_{1,3}\cap\mathcal
D_\infty.$ Since the radius of $\mathcal D_R$ was chosen initially
to be less than one, the contour $\partial\mathcal D_R$ crosses
the interval $[0,1].$ The integration variable $k$ in
\eref{detaposik} lies in this interval and, therefore, the
estimate \eref{schi} can not be applied everywhere on the
integration contour. This problem is circumvented by the analytic
continuation of equation \eref{schi} from the disk $\mathcal D_R$
into a bigger region $\Omega_R$ containing the interval $[0,1],$
as shown in figure \ref{Fig:OmegaRcontinuation}.
The matrices $Q_{R}(k)$ and $\tilde S_R(k)$ are given in
$\Omega_R$ by the same formulas as in $\mathcal D_R:$ the matrix
$Q_R$ by \eref{AnsaR}, \eref{E} and \eref{phientries}; the matrix
$\tilde S_R$ by \eref{jrRinverse}. Using \eref{f=ye} and
\eref{Y=Unu} one gets the approximation to $\vec f(k)$
\begin{equation}
\vec f = \tilde S_R\vec g\, [I+\mathcal{O}(x^{-1-\epsilon})]
\label{f=Sg}
\end{equation}
uniform for $k\in[0,1].$ The function $\vec g$ is defined by the
equation
\begin{equation}
\vec g(k)=\left(
\begin{array}{c}
g_+(k) \\
g_-(k)
\end{array} \right)=
[Q_R(k)]_+\nu_1(k) \vec e(k) \qquad k\in[0,1]. \label{g=PhiRnue}
\end{equation}
The matrix $\nu_1$ entering \eref{g=PhiRnue} is given by equation
\eref{nu1def}, the function $\vec e$ by \eref{vecfe} and
\eref{epm0}, the matrix $Q_R(k)$ by \eref{phientries},
\eref{AnsaR} and \eref{E}. With the help of the identity
\cite{BE-53}
\begin{equation}
\fl {}_1F_1(a, b; w)=\frac{\Gamma(b)}{\Gamma(b-a)}\rme^{\rmi \pi a
\lambda }\Psi(a,b;w)+ \frac{\Gamma(b)}{\Gamma(a)} \rme^{\rmi \pi
(a-b)\lambda} \rme^w \Psi(b-a, b; -w) \label{F11iden}
\end{equation}
where $ \lambda={\rm sgn \ Im}w,$ one gets for \eref{g=PhiRnue}
\begin{equation}
\fl\eqalign{ g_{+}(k)= -\frac{\sqrt{\pi} \rme^{\frac{\rmi
\eta}{4}}} {\Gamma\left( -\frac{\eta}{2 \pi}    \right)}
\rme^{-\frac{\rmi k x}{2}+\rmi x} x^{-\frac{\eta}{2\pi}}
(k+1)^{-\frac \eta {2 \pi }} {}_1F_1\left[\frac{\eta}{2\pi}+1, 1;
\rmi x (k-1)  \right]
\\
g_{-}(k)=\frac{\sqrt{\pi} \rme^{-\frac{\rmi
\eta}{4}}}{\Gamma\left(\frac{\eta}{2\pi}\right)\sin\frac{\eta}{2}
} \rme^{- \frac{\rmi k x}{2}}
 x^{\frac{\eta}{2\pi}} (k+1)^{\frac \eta {2 \pi }}
{}_1F_1\left[\frac{\eta}{2\pi}, 1; \rmi x (k-1)  \right].}
\label{gexplicit}
\end{equation}

Using the uniform estimate \eref{f=Sg} and the formulas
\eref{gexplicit} one can derive the following estimate for
\eref{detaposik}
\begin{equation}
\fl\partial_\eta\ln\det(\hat I +\hat V)= \frac{2}{1-\rme^{-\rmi
\eta}} \int_{0}^{1} \rmd k [\vec g(k)]^{T} \sigma_2 \partial_k
\vec g(k)[1+\mathcal{O}(x^{-1-\epsilon})]. \label{detagdkg}
\end{equation}
Performing the integral in \eref{detagdkg} asymptotically one gets
\begin{equation}
\partial_\eta\ln\det(\hat I +\hat V)= \frac{\rmi x}{\pi}- \frac{\eta}{\pi^2} \ln x
+C_1(\eta)+\mathcal O\left(x^{-\epsilon} \right) \label{detagdkg2}
\end{equation}
where
\begin{equation}
C_1(\eta)=\frac{\eta}{2\pi^2}\left[\psi\left(\frac{\eta}{2\pi}+1\right)+
\psi\left(-\frac{\eta}{2 \pi}+1\right)- 2(\ln 2+1) \right]
\label{C1}
\end{equation}
and $\psi(w)$ is the digamma function,
\begin{equation}
\psi(w)=\frac\rmd{\rmd w}\ln\Gamma(w).
\end{equation}
Integrating equation \eref{detagdkg2} over $\eta,$ taking into
account the initial condition \eref{inicondeta} and comparing the
resulting expression with \eref{logdet} one gets
\begin{equation}
\fl C(\eta)= \frac{\eta}{\pi}\ln\left[
\frac{\Gamma\left(1+\frac{\eta}{2\pi}\right)} {\Gamma\left(1-
\frac{\eta}{2\pi}\right)} \right] - \frac{\eta^2}{2 \pi^2} (\ln
2+1) - \frac{1}{\pi} \int_0^{\eta} \rmd \lambda \ln\left[
\frac{\Gamma\left(1+\frac{\lambda}{2\pi}\right)} {\Gamma\left(1-
\frac{\lambda}{2\pi}\right)} \right]. \label{Cdef}
\end{equation}

\subsection{Equal-time correlation functions: the results \label{answer0}}
In this subsection we give the answer for the correlation
functions $G_h(x,0)$ and $G_e(x,0)$ in the large $x$ limit.

The correlation function $G_h(x,0)$ is given by the formula
\eref{Gx0res}. The asymptotics of $B_{--}$ is given by
\eref{Bmmas}; the asymptotics of $\det(\hat I+\hat V)$ by
\eref{logdet} with the constant $C(\eta)$ given by \eref{Cdef}.
Recall that $z=\exp(\rmi \eta),$ therefore the point $z=1/2$
corresponds to $\eta=\rmi \ln 2.$ The final answer for $G_h(x,0)$
is
\begin{equation}
G_h(x,0)=\Xi x^{-\alpha}{\sin\left(x-x_0 -\frac{\ln 2}{\pi} \ln x
\right)}\exp\left(-\frac{\ln 2}{\pi} x\right)\label{G0result}
\end{equation}
where the anomalous exponent $\alpha$ is given by
\begin{equation}
\alpha=1-\frac{1}{2}\left(\frac{\ln2}{\pi}\right)^2
\label{anomdim}
\end{equation}
the phase shift $x_0$ by
\begin{equation}
x_0=\frac{(\ln 2)^2}{\pi}-2 \rm{Im}\left[ \ln\Gamma\left(
\frac{\rmi \ln 2}{2\pi}\right)\right] \label{x0def}
\end{equation}
and the constant $\Xi$ by
\begin{equation}
\Xi=-{4\pi \sqrt 2} \exp\left\{C(\rmi \ln 2) - 2 {\rm Re} \left[
\ln \Gamma\left( \frac{\rmi \ln 2}{2 \pi}\right) \right] \right\}
\label{Xidef}
\end{equation}
where $C(\eta)$ is defined by equation \eref{Cdef}. The relative
correction to \eref{G0result} is of the order of $x^{-1}.$ Note
that equations \eref{G0result} and \eref{anomdim} were first
derived in paper \cite{BL-87}. The phase shift \eref{x0def} and
the constant \eref{Xidef} were calculated in paper \cite{B-91}.

The asymptotics of the function $G_e(x,0)$ is obtained from
equation \eref{G0result} by using equation \eref{GeGhdelta}.
Recall that in the derivation of equation \eref{G0result} we
assumed that $x$ is positive. The result for negative $x$ follows
from equation \eref{xtominusx}.


\section{Riemann-Hilbert problem at $t\ne0$. Space-like region \label{section:RHPtime}}

In this section we find the asymptotic solution of the
Riemann-Hilbert problem \eref{RHPtime} for $|x|>2| t|.$ We assume
that $x>0$ and $t>0.$ We will consider the large $x$ asymptotics
for a fixed value of
\begin{equation}
k_s=\frac{x}{2 t}. \label{ksdef}
\end{equation}

In subsection \ref{reformulation} we reformulate the RH problem
\eref{RHPtime} so as to make it similar to the RH problem
\eref{RHPx}. In subsections \ref{lenstsl} and \ref{lenstsl2} we
repeat the construction made for the $t=0$ case in subsections
\ref{deformed1} and \ref{deformed2}, respectively. In subsection
\ref{QRLt} we give an explicit solution to the jump relation
(\ref{RHPtZ}.b) in a vicinity of the end points $k=\pm 1$ of the
contour $\Sigma_0.$ In subsection \ref{Sast} we perform the
matching procedure analogous to that of subsection
\ref{perturthe}. In subsection \ref{approxt} we summarize the
results of this section.

\subsection{Reformulation of the Riemann-Hilbert problem \eref{RHPtime} \label{reformulation}}

In this subsection we reformulate the RH problem \eref{RHPtime} so
as to make it similar to the RH problem \eref{RHPx}.

\begin{figure}
\centering
\includegraphics[width=0.5\textwidth]{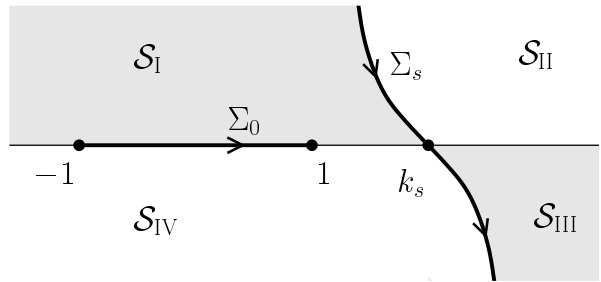}
\caption{The conjugation contour $\Sigma_0\cup\Sigma_s$ for the RH
problem \eref{RHPtZ}. The sectors $\mathcal S_\mathrm{I}$ and
$\mathcal S_\mathrm{III}$ are shaded in gray.}
\label{Fig:tcontour}
\end{figure}

Represent $Y(k)$ as a product of two matrices
\begin{equation}
Y(k)=Z(k) V(k) \label{Y=ZV}
\end{equation}
and define $V(k)$ by
\begin{equation}
V(k)=\left (
\begin{array}{cc}
1 & - 2 \sin^2 \frac\eta 2 \left[E(k)- \rmi \rme^{-\tau(k)} \varrho(k)\right] \\
0 &1
\end{array} \right)
\label{V}
\end{equation}
with $E(k)$ given by equation \eref{Edef} and $\varrho(k)$ being a
piecewise-constant function
\begin{equation}
\varrho(k)=\left  \{
\begin{array}{rl}
1 &  \quad k\in \mathcal S_{\rm I} \\
-1 &\quad k\in \mathcal S_{\rm III} \\
0&  \quad {\rm otherwise}
\end{array}
\right. .\label{rho}
\end{equation}
The sectors $\mathcal S_{\rm I}$ and $\mathcal S_{\rm III}$ are
shown in figure \ref{Fig:tcontour}.
The oriented contour $\Sigma_s$ separating the sector $\mathcal
S_{\rm I}$ from $\mathcal S_{\rm II}$ and the sector $\mathcal
S_{\rm III}$ from $\mathcal S_{\rm IV}$ is chosen so that (a) it
passes through the point $k_s$ (b) the real part of $\tau(k)$ is
positive for all $k\in\Sigma_s$  except $k=k_s$
\begin{equation}
{\rm Re}\tau(k) >0 \qquad k\in \Sigma_s \setminus\{k_s\}
\label{decaycondition1}
\end{equation}
and
\begin{equation}
{\rm Re}\tau(k)\to+\infty \qquad\mathrm{as}\qquad k\to\infty\qquad k\in\Sigma_s.
\label{decaycondition2}
\end{equation}
It is easy to see that $V(k)$ is analytic everywhere in the
complex $k$-plane except the contour $\Sigma_s,$ where it
satisfies the jump relation
\begin{equation}
V_{+}(k)=[\mu_s(k)]^{-1}V_{-}(k) \qquad k\in \Sigma_s
\label{museq}
\end{equation}
with
\begin{equation}
\mu_s(k)=\left(
\begin{array}{cc}
1& 2 \rmi \rme^{-\tau(k)} \sin^2\frac{\eta}{2} \\
0&1
\end{array} \right).
\label{musdef}
\end{equation}
Note that the condition \eref{decaycondition2} implies
\begin{equation}
V(k)\to I \qquad {\rm as } \qquad k\to \infty. \label{Vnormalized}
\end{equation}

With $V(k)$ and $\Sigma_s$ defined above,
one gets the following RH problem for $Z(k)$
\begin{eqnarray}
\eqalign{\mbox{(a)}\quad& Z(k) \mbox{ is  analytic  in } {\mathbb C} \setminus (\Sigma_0\cup\Sigma_s) \\
\mbox{(b)}\quad& Z_{+}(k)=Z_{-}(k)\mu_{i}(k) \qquad  k \in \Sigma_{i} \qquad i=0,s \\
\mbox{(c)}\quad& Z(k)\to I \mbox{ as } k\to \infty} \label{RHPtZ}
\end{eqnarray}
where the contours $\Sigma_0$ and $\Sigma_s$ are shown in figure
\ref{Fig:tcontour}. Since $Y(k)$ is analytic across $\Sigma_s,$
the matrix $\mu_s$ should satisfy \eref{museq} and, therefore, is
given by \eref{musdef}. The matrix $\mu_0$ can be obtained
comparing the jump relations (\ref{RHPtime}.b) and (\ref{RHPtZ}.b)
\begin{equation}
\mu_{0}(k)=V(k)\nu_0(k;x,t)[V(k)]^{-1} \qquad k\in\Sigma_0.
\label{muV}
\end{equation}
Representing \eref{Gcondef} in the form
\begin{equation}
\fl \nu_0(k;x,t)=\left(
\begin{array}{cc}
1 & 2 \sin^2\frac{\eta}{2} E_{-}(k) \\ 0 &1
\end{array}
\right)
\left(
\begin{array}{cc}
\rme^{\rmi\eta}&0\\
2\rmi\rme^{\tau(k)}&\rme^{-\rmi \eta}
\end{array}
\right)
\left(
\begin{array}{cc}
1 & -2 \sin^2\frac{\eta}{2}E_{+}(k) \\ 0 &1
\end{array}
\right) \label{nu0defr2}
\end{equation}
with $E_{\pm}$ given by \eref{Edef} and \eref{Epmprop} one gets
\begin{equation}
\mu_0(k)=\left(
\begin{array}{cc}
\rme^{\rmi \eta}& -2 \rmi \rme^{-\tau(k)}\rme^{\rmi \eta}\sin^2\frac{\eta}2\\
2 \rmi \rme^{ \tau(k) }&2-\rme^{\rmi \eta}
\end{array}\right) .
\label{mu0def}
\end{equation}
We see that the matrix $\mu_0$ is different  from the matrix
$\nu_0,$ defined by \eref{nu0def}, only in that $\rmi k x $ in the
latter is replaced by $-\tau(k)$ in the former. The matrix $\mu_0$
obeys the involution
\begin{equation}
P\mu_0(k;x,t)P^{-1}=[\mu_0(-k;x,-t)]^{-1} \label{mu0involution}
\end{equation}
with the matrix $P$ defined by equation \eref{Pdef}. This
involution is a natural generalization of the involution
\eref{involution}.

\subsection{Riemann-Hilbert problem on the contour $\Sigma_{\rm
lens}\cup \Sigma_{s}$ \label{lenstsl}}

In this subsection an RH problem equivalent to the RH problem
\eref{RHPtZ} is formulated on the conjugation contour $\Sigma_{\rm
lens}\cup \Sigma_{s}.$ The procedure will resemble that of
subsection \ref{deformed1}.

\begin{figure}
\centering
\includegraphics[width=0.5\textwidth]{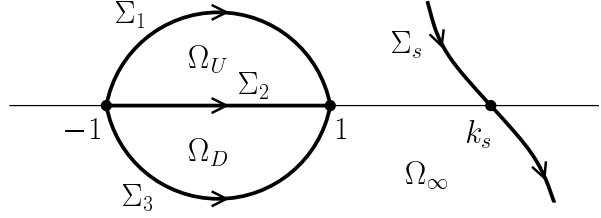}
\caption{The conjugation contour for the RH problem \eref{RHPtU}.
The contour consists of the lens-shaped contour
$\Sigma_\mathrm{lens}=\Sigma_1\cup\Sigma_2\cup\Sigma_3$ and the
contour $\Sigma_s.$} \label{Fig:tlens}
\end{figure}

Represent $Z(k)$ as a product of two matrices
\begin{equation}
Z(k)=U(k)L(k) \label{Z=UL}
\end{equation}
and define $L(k)$ by the formula
\begin{equation}
L(k) =\left\{
\begin{array}{ll}
I &\qquad k\in\Omega_{\infty} \\
\mu_1(k) &\qquad k\in\Omega_{U}\\
\left[\mu_{3}(k)\right]^{-1} &\qquad k\in\Omega_{D}
\end{array}
\right. .\label{Ldeft}
\end{equation}
The lens-shaped contour
$\Sigma_\mathrm{lens}=\Sigma_1\cup\Sigma_2\cup\Sigma_3$ is the
same as in subsection \ref{deformed1}. It is shown in
figure~\ref{Fig:tlens}.
The matrices $\mu_1$ and $\mu_3$ are defined as
\begin{eqnarray}
\mu_1(k)&=
\left(\begin{array}{cc}1&-2\rmi\rme^{-\tau(k)}\sin^2\frac{\eta}2\\
0 &1
\end{array}\right) \vspace{20pt} \label{mu1def}
\\
\mu_3(k)&=\left(\begin{array}{cc} 1& 0\\ 2 \rmi \rme^{\tau(k)}
\rme^{-\rmi \eta}&1 \end{array}\right). \label{mu3def}
\end{eqnarray}
For the conjugation matrix \eref{mu0def} the following
representation can be written
\begin{equation}
\mu_0=\mu_3 \mu_2 \mu_1 \label{trimu}
\end{equation}
where the matrix
\begin{equation}
\mu_2=\left(\begin{array}{cc} \rme^{\rmi \eta}& 0\\
0&\rme^{-\rmi \eta} \end{array}\right) \label{mu2def}
\end{equation}
coincides with the matrix $\nu_2$ defined by equation
\eref{nu2def}. Note that
\begin{equation}
P\mu_1(k; x,t)P^{-1}=[\mu_3(-k;x,-t)]^{-1}\qquad P\mu_2P^{-1}=\mu_2^{-1}
\label{mu13inv}
\end{equation}
where the matrix $P$ is defined by \eref{Pdef}.

It follows from equations \eref{RHPtZ}, \eref{Z=UL}, \eref{Ldeft}
and \eref{trimu} that $U(k)$ solves the following RH problem
\begin{eqnarray}
\eqalign{\mbox{(a)}\quad& U(k) \mbox{ is analytic in }
{\mathbb C} \setminus  (\Sigma_\mathrm{lens}\cup\Sigma_s)\\
\mbox{(b)}\quad& U_{+}(k)=U_{-}(k)\mu_{i}(k) \qquad  k \in \Sigma_{i} \qquad i=1,2,3,s \\
\mbox{(c)}\quad& U(k)\to I \mbox{ as } k\to \infty} \label{RHPtU}
\end{eqnarray}
where the contours $\Sigma_i$ are shown in figure \ref{Fig:tlens}.
We will calculate the large $x$ asymptotics of $U(k);$ the
asymptotics of $Z(k)$ will then follow from \eref{Z=UL}.

\subsection{Factorization of the Riemann-Hilbert problem \eref{RHPtU} \label{lenstsl2}}
In this subsection we present the scheme of the analysis of the RH
problem \eref{RHPtU} in the large $x$ limit. The scheme is similar
to the one given in subsection \ref{deformed2}.

\begin{figure} \centering
\includegraphics[width=0.5\textwidth]{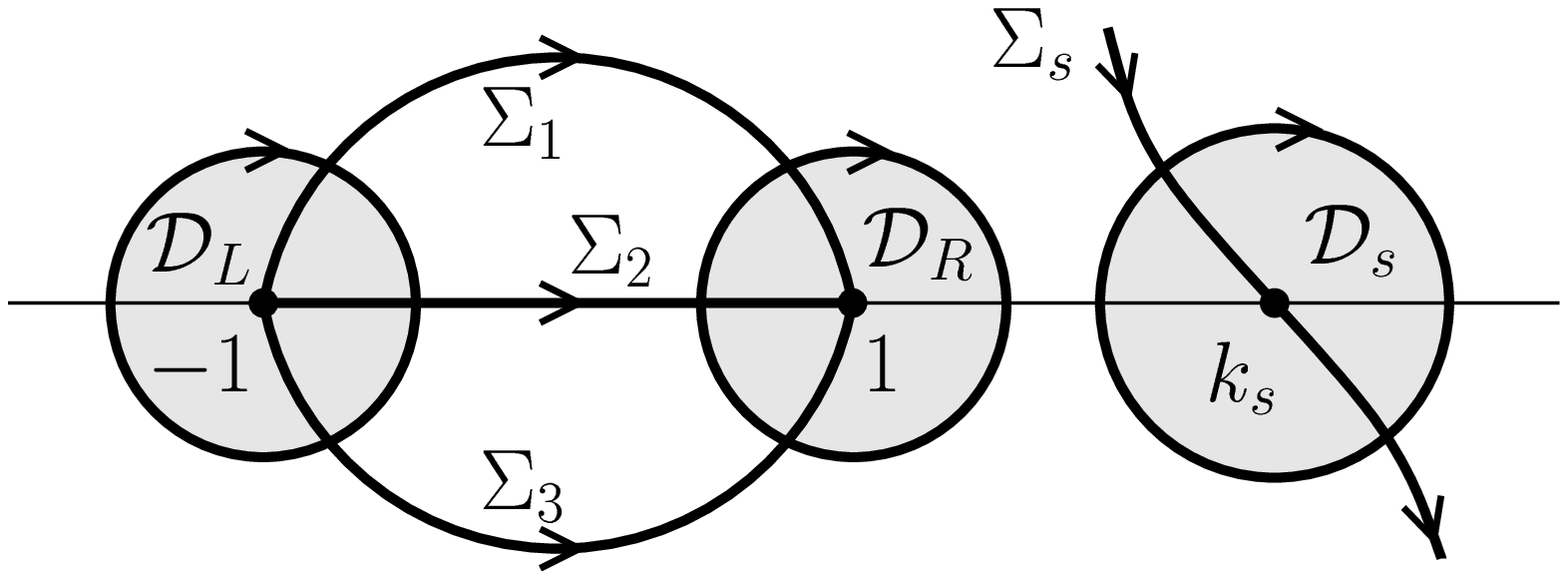}
\caption{The contour used in the factorization of the RH problem
\eref{RHPtU}. The disks $\mathcal D_R,$ $\mathcal D_L$ and
$\mathcal D_s$ are shaded in gray.} \label{Fig:tsplit}
\end{figure}

Like the $t=0$ case, the jump matrix $\mu_1$ (respectively,
$\mu_3$) goes to the identity matrix, as $x\to\infty,$ everywhere
on the contour $\Sigma_1$ (respectively, $\Sigma_3$) except at the
points $k=\pm1.$ The jump matrix $\mu_s$ goes to the identity
matrix everywhere on $\Sigma_s$ except at the point $k=k_s.$
Therefore, to construct a uniform approximation to $U(k)$ we use
the same scheme as in subsection \ref{deformed2}. Introduce three
disks $\mathcal D_R,$ $\mathcal D_L$ and $\mathcal D_s$ as shown
in figure~\ref{Fig:tsplit}.
Denote the domain $\mathbb C\setminus(\mathcal D_R\cup\mathcal
D_L\cup\mathcal D_s)$ as $\mathcal D_\infty.$ Represent the
solution $U(k)$ of the RH problem \eref{RHPtU} as a product of
three matrices
\begin{equation}
U=KSQ. \label{U=KSQt}
\end{equation}
The matrix $Q(k)$ is given by the formula
\begin{equation}
Q(k)=
\left\{
\begin{array}{ll} Q_{R,L}(k)&\qquad k\in\mathcal D_{R,L}\\
Q_s(k)&\qquad k\in\mathcal D_s\\
Q_\infty(k)&\qquad k\in\mathcal D_\infty
\end{array}
\right. .\label{Qdeft}
\end{equation}
The matrix $Q_\infty$ is given by equation \eref{Ndef}. The matrix
$Q_s$ is defined in $\mathcal D_s$ and satisfies
\begin{equation}
\left[Q_s(k)\right]_+= \left[Q_s(k)\right]_{-}\mu_s(k) \qquad k
\in \Sigma_s\cap\mathcal D_s .\label{Qjumpts}
\end{equation}
We choose the following solution of the jump relation
\eref{Qjumpts}:
\begin{equation}
Q_s(k)= Q_\infty(k)\left(
\begin{array}{cc}
1& 2\sin^2\frac{\eta}{2} \int_{\Sigma_s} \frac{\rmd p}{2\pi}
\frac{\rme^{-\tau(p)}}{p-k} \\
0&1
\end{array}\right)
\label{Qs}
\end{equation}
where the integral runs along the contour $\Sigma_s$ shown in
figure \ref{Fig:tsplit}. The matrices $Q_R$ and $Q_L$ are defined
in the disks $\mathcal D_R$ and $\mathcal D_L,$ respectively, and
satisfy there the jump relations
\begin{equation}
\left[Q_R(k)\right]_+= \left[Q_R(k)\right]_{-}\mu_i(k) \qquad k
\in \Sigma_i\cap\mathcal D_R \qquad i=1,2,3 \label{QjumptR}
\end{equation}
and
\begin{equation}
\left[Q_L(k)\right]_+= \left[Q_L(k)\right]_{-}\mu_i(k) \qquad k
\in \Sigma_i\cap\mathcal D_L \qquad i=1,2,3. \label{QjumptL}
\end{equation}
We give solutions to the jump relations \eref{QjumptR} and
\eref{QjumptL} in subsection \ref{QRLt}.

The matrix $S$ is defined as a solution of the RH problem
\begin{eqnarray}
\eqalign{\mbox{(a)}\quad& S(k) \mbox{ is analytic in }
\mathbb{C}\setminus(\partial \mathcal D_R\cup\partial\mathcal D_L\cup \partial \mathcal D_s) \\
\mbox{(b)}\quad& S_{+}(k)=S_{-}(k)\theta_{\alpha } (k) \qquad  k \in \partial \mathcal D_\alpha \qquad \alpha=R,L,s\\
\mbox{(c)}\quad& S(k)\to I \mbox{ as } k\to \infty} \label{RHPst}
\end{eqnarray}
with the clockwise oriented conjugation contours $\partial\mathcal
D_{R,L,s}$ shown in figure~\ref{Fig:tsplit} and the conjugation
matrices $\theta_{R,L,s}$ defined by
\begin{equation}
\theta_\alpha(k)=Q_\alpha (k) \left[Q_\infty(k)\right]^{-1} \qquad
k\in \partial \mathcal D_\alpha \qquad \alpha=R,L,s.
\label{thetaRLsdef}
\end{equation}
We solve the RH problem \eref{RHPst} in the large $x$ limit in
subsection \ref{Sast}.

The matrix $K$ in the decomposition \eref{U=KSQt} is constructed
in complete analogy with the matrix $K$ considered in subsection
\ref{deformed2}.
\subsection{Exact solutions to the jump relations
\eref{QjumptR} and \eref{QjumptL} \label{QRLt}}

In this subsection we solve explicitly the jump relations
\eref{QjumptR} and \eref{QjumptL} by mapping them on the jump
relations \eref{QRjump} and \eref{QLjump}, solved in subsection
\ref{explicit}.

\begin{figure}
\centering
\includegraphics[width=0.6\textwidth]{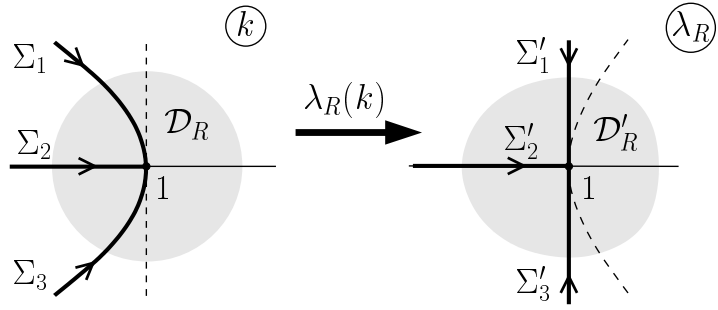}
\caption{Conformal map $\lambda_R.$ The disk $\mathcal D_R$ and
its image $\mathcal D_R^\prime$ are shaded in gray. The
conjugation contours $\Sigma_1$ and $\Sigma_3$ are chosen so that
their images $\Sigma_1^\prime$ and $\Sigma_3^\prime$ are parallel
to the imaginary axis.}
 \label{Fig:cmap}
\end{figure}

Introduce the `light cone' variables $x_R$ and $x_L$
\begin{equation}
x_{R}=x-2t \qquad x_{L}= x+2t \label{xRLdef}.
\end{equation}
Consider the jump relation \eref{QjumptR} and make the following
conformal map
\begin{equation}
\lambda_R(k)=1-\frac{\tau(k)-\tau(1)}{\rmi x_R} \qquad
k\in\mathcal D_R \label{zdef}
\end{equation}
of the disk $\mathcal D_R$ in the complex $k$-plane onto a
vicinity $\mathcal D^\prime_R$ of the point $\lambda_R=1$ in the
$\lambda_R$-plane, as shown in figure \ref{Fig:cmap}.
Introducing a $k$-independent matrix $T$ by the formula
\begin{equation}
T= \left(
\begin{array}{cc}
\rme^{-{\rmi t}/{2}} & 0 \\
0 & \rme^{{\rmi t}/{2}}
\end{array}
\right) \label{TRLdef}
\end{equation}
one can easily check that
\begin{equation}
\mu_i(k)=T^{-1}\nu_i[\lambda_R(k);x_R]T \qquad k\in\mathcal D_R
\qquad i=1,2,3. \label{TmuT}
\end{equation}
Here the matrices $\mu_1,$ $\mu_3$ and $\mu_2$ are defined by the
formulas \eref{mu1def}, \eref{mu3def} and \eref{mu2def},
respectively; the matrices $\nu_1,$ $\nu_3$ and $\nu_2$ are
defined by the formulas \eref{nu1def}, \eref{nu3def} and
\eref{nu2def}, respectively. Choose the shape of the contours
$\Sigma_1$ and $\Sigma_3$ in the $k$-plane so that their images
$\Sigma^\prime_1$ and $\Sigma^\prime_3$ in the $\lambda_R$-plane
are parallel to the imaginary axis, as shown in
figure~\ref{Fig:cmap}. Write equation \eref{PhiRjump} with $k$
replaced by $\lambda_R$ and $x$ replaced by $x_R:$
\begin{equation}
\fl\left[\Phi_R(\lambda_R;x_R)\right]_{+}=\left[\Phi_R(\lambda_R;x_R)\right]_{-}
\nu_i(\lambda_R;x_R) \qquad \lambda_R\in \Sigma^\prime_i \qquad
i=1,2,3 \label{Phijumpt}
\end{equation}
where $\Phi_R$ is defined by \eref{phientries}. By virtue of
\eref{TmuT} the jump relation \eref{Phijumpt} is equivalent to
\begin{equation}
\fl\left\{\Phi_R[\lambda_R(k);x_R]T\right\}_{+}=
\left\{\Phi_R[\lambda_R(k);x_R]T\right\}_{-}\mu_i(k) \qquad
k\in\Sigma_i \qquad i=1,2,3.
\end{equation}
One sees that the matrix $\Phi_R[\lambda_R(k);x_R]T$ solves the
jump relation \eref{QjumptR} in the disk $\mathcal D_R.$ Now we
are in a position to write down an ansatz for $Q_R(k):$
\begin{equation}
Q_R(k)=E_R(k)\Phi_R[\lambda_R(k);x_R]T \label{QRt}
\end{equation}
where
\begin{equation}
E_R(k)=Q_\infty(k)\left(
\begin{array}{cc}
[\lambda_R(k)-1]^{-\frac\eta{2\pi}} & 0 \\
0 & [\lambda_R(k)-1]^{\frac\eta{2\pi}}
\end{array}
\right) T^{-1}. \label{ERtdef}
\end{equation}
With this choice of $E_{R}(k)$ the matrix $Q_R(k)$ goes to
$Q_{\infty}(k),$ as $x\to\infty,$ for any $\epsilon\ne0.$

Taking into account the involution \eref{mu13inv} choose the following
solution to the jump relation \eref{QjumptL}
\begin{equation}
Q_L(k;x,t)=PQ_R(-k;x,-t)P^{-1} \label{QLt} \qquad k \in \mathcal D_L
\end{equation}
where $Q_R$ is given by formulas \eref{QRt} and \eref{ERtdef}.

\subsection{Solution to the Riemann-Hilbert problem
\eref{RHPst} in the large $x$ limit \label{Sast}}

In this subsection we solve the RH problem \eref{RHPst} in the
large $x$ limit. The procedure will resemble that of
subsection~\ref{perturthe}.

Consider the conjugation matrices $\theta_R$ and $\theta_L$
defined by equation \eref{thetaRLsdef}. They satisfy the involution
\begin{equation}
\theta_L(k;x,t)=P\theta_R(-k;x,-t)P^{-1} \qquad k \in \partial \mathcal D_L \label{thetainvt}
\end{equation}
as can be shown using the involutions \eref{Qinv} and \eref{QLt}.
Substituting equations \eref{QRt} and \eref{ERtdef} into equation
\eref{thetaRLsdef} and using the results of subsection
\ref{explicit} one gets the following approximation of the matrix
$\theta_R$ by a matrix $\tilde \theta_R$
\begin{equation}
\theta_{R}=\tilde \theta_{R}[I+\mathcal O
\left(x^{-1-\epsilon}\right)]
\label{thetaRassl}
\end{equation}
uniform for $k\in\partial\mathcal D_R.$ Here
\begin{equation}
\tilde \theta_{R}(k)=I+\frac{\upsilon_{R}(k)}{x(k-1)}\qquad k\in
\partial \mathcal D_R\label{thetaRast}.
\end{equation}
The matrix $\upsilon_R$ is given by
\begin{equation}
\fl \upsilon_{R}(k)=\left(
\begin{array}{cc}
-\rmi \left(\frac{\eta}{2 \pi}\right)^2 \frac{2 k_s}{2 k_s-k-1} &
a_R \left[x (k+1) \right]^{-\frac{\eta}{\pi}} \left(\frac{2 k_s}
{2 k_s-k-1}\right)^{1+\frac{\eta}{\pi}}
\\
b_R \left[x(k+1) \right]^{\frac{\eta}{\pi}}\left(\frac{2 k_s}
{2k_s-k-1}\right)^{1-\frac{\eta}{\pi}} & \rmi \left(\frac{\eta}
{2\pi}\right)^2 \frac{2 k_s}{2 k_s-k-1}
\end{array} \right)
\label{upsilonRt}
\end{equation}
where
\begin{equation}
a_R= -\frac{\rmi\pi\rme^{\rmi\frac\eta2}\rme^{\rmi (x-t)}}
{\Gamma^2\left(-\frac\eta{2\pi}\right)} \qquad
b_R=-\frac{\rmi\pi\rme^{-\rmi\frac\eta2}\rme^{-\rmi (x-t)}}
{\Gamma^2\left(\frac\eta{2\pi}\right)\sin^2\frac\eta2}.
\end{equation}
It follows from the involution \eref{thetainvt} that the matrix
$\theta_L$ is approximated by a matrix $\tilde\theta_L$
\begin{equation}
\theta_{L}=\tilde \theta_{L}[I+\mathcal O
\left(x^{-1-\epsilon}\right)] \label{thetaLassl}
\end{equation}
uniformly for $k\in\partial\mathcal D_L,$ where
\begin{equation}
\tilde\theta_L(k;x,t)=P\tilde\theta_R(-k;x,-t)P^{-1}=I
+\frac{\upsilon_L(k)}{x(k+1)} \qquad k\in\partial\mathcal D_L
\label{thetaLast}
\end{equation}
and
\begin{equation}
\upsilon_L(k;x,t)= -P\upsilon_R(-k;x,-t)P^{-1} \label{upsilonLsl}.
\end{equation}
The explicit expression for $\upsilon_L$ reads
\begin{eqnarray}
\fl \upsilon_{L}(k)=\left(
\begin{array}{cc}
-\rmi \left(\frac{\eta}{2 \pi}\right)^2 \frac{2 k_s}{2 k_s-k+1} &
a_L \left[x (1-k) \right]^{\frac{\eta}{\pi}} \left(\frac{2 k_s}
{2k_s-k+1}\right)^{1-\frac{\eta}{\pi}}
\\
b_L \left[x (1-k) \right]^{-\frac{\eta}{\pi}} \left(\frac{2 k_s}
{2 k_s-k+1}\right)^{1+\frac{\eta}{\pi}} & \rmi \left(\frac{\eta}
{2\pi}\right)^2 \frac{2 k_s}{2 k_s-k+1}
\end{array} \right)
\label{upsilonLt}
\end{eqnarray}
where
\begin{eqnarray}
a_L = \frac{\rmi\pi\rme^{\rmi\frac\eta2}\rme^{-\rmi (x+t)}}
{\Gamma^2\left(\frac\eta{2\pi}\right)} \qquad & b_L
=\frac{\rmi\pi\rme^{-\rmi\frac\eta2}\rme^{\rmi (x+t)}}
{\Gamma^2\left(-\frac\eta{2\pi}\right)\sin^2\frac\eta2}.
\end{eqnarray}

To calculate the conjugation matrix $\theta_s$ in the large $x$
limit use the large $x$ asymptotics of the integral in \eref{Qs}
\begin{equation}
\int_{\Sigma_s} \frac{\rmd p}{2\pi}
\frac{\rme^{-\tau(p)}}{p-k}=-\frac{G_0}{k-k_s}+\mathcal
O(x^{-3/2}) \label{qas}
\end{equation}
where $G_0$ is defined by equation \eref{Gvac}. Note that the
leading term in \eref{qas} is of the order of $x^{-1/2}.$
Substituting equation \eref{qas} into equation \eref{Qs} and using
equation \eref{thetaRLsdef} one gets
\begin{equation}
\theta_s=\tilde \theta_s[I +\mathcal O (x^{-3/2})]
\label{thetastilde}
\end{equation}
uniformly for $k\in\partial\mathcal D_s,$ where
\begin{equation}
\tilde\theta_s(k) =I+\frac{\upsilon_s(k)}{k-k_s} \qquad
k\in\partial\mathcal D_s \label{thetasas} \label{tildetetaSsl}
\end{equation}
and
\begin{equation}
\upsilon_s(k)=
\left(
\begin{array}{cc}
0 &-2G_0\sin^2\frac{\eta}{2} \left(\frac{ k-1}{k+1}\right)^{\frac{\eta}{\pi}} \\
0& 0
\end{array}
\right).
\end{equation}

Consider the RH problem
\begin{eqnarray}
\eqalign{\mbox{(a)}\quad& \tilde S(k) \mbox{ is analytic in }
\mathbb{C}\setminus(\partial \mathcal D_R\cup\partial\mathcal D_L \cup \partial \mathcal  D_s) \\
\mbox{(b)}\quad& \tilde S_{+}(k)=\tilde S_{-}(k)\tilde \theta_{\alpha } (k) \qquad  k \in \partial \mathcal D_{\alpha}
\qquad \alpha = R, L, s \\
\mbox{(c)}\quad& \tilde S(k)\to I \mbox{ as } k\to \infty}
\label{RHPstildet}
\end{eqnarray}
with the jump matrices $\tilde\theta_R,$ $\tilde\theta_L$ and
$\tilde\theta_s$ defined by \eref{thetaRast}, \eref{thetaLast} and
\eref{thetasas}, respectively. In complete analogy with
equation \eref{Stilde} the solution $\tilde S(k)$ of the RH
problem \eref{RHPstildet} approximates the solution $S(k)$ of the
RH problem \eref{RHPst}
\begin{equation}
S(k)=\tilde S(k)[I+\mathcal O(x^{-1-\epsilon})] \label{StStilde}
\end{equation}
uniformly outside a vicinity of the conjugation contours
$\partial\mathcal D_R,$ $\partial\mathcal D_L$ and
$\partial\mathcal D_s.$

To solve the RH problem \eref{RHPstildet} represent $\tilde S(k)$
as a product of two matrices
\begin{equation}
\tilde S(k)= M(k) N(k) \label{S=MN}
\end{equation}
where the matrix $N(k)$ is the solution to the RH problem
\begin{eqnarray}
\eqalign{
\mbox{(a)}\quad&  N(k) \mbox{ is analytic in }
\mathbb{C}\setminus(\partial \mathcal D_R\cup\partial\mathcal D_L) \\
\mbox{(b)}\quad& N_{+}(k)=N_{-}(k)\tilde\theta_{R,L}(k) \qquad k
\in\partial\mathcal D_{R,L} \\
\mbox{(c)}\quad&  N(k)\to I \mbox{ as } k\to\infty} \label{RHPNt}
\end{eqnarray}
with the jump matrices $\tilde\theta_R$ and $\tilde\theta_L$ given
by \eref{thetaRast} and \eref{thetaLast}, respectively. Comparing
\eref{RHPstildet} and \eref{RHPNt} one sees that the matrix $M(k)$
solves the following RH problem
\begin{eqnarray}
\eqalign{\mbox{(a)}\quad& M(k) \mbox{ is analytic in }
\mathbb{C}\setminus\partial\mathcal D_s \\
\mbox{(b)}\quad& M_{+}(k)= M_{-}(k)N(k)\tilde\theta_s (k)
N^{-1}(k) \qquad k\in\partial\mathcal D_{s}
\\
\mbox{(c)}\quad& M(k)\to I \mbox{ as } k\to \infty} \label{RHPMt}
\end{eqnarray}
where the jump matrix $\tilde\theta_s$ is defined by equation \eref{thetasas}.

Let us solve the RH problem \eref{RHPNt} first. Since the jump
matrices \eref{thetaRast} and \eref{thetaLast} have the same
analytic properties in the complex $k$-plane as the matrices
\eref{thetaRas} and \eref{thetaLas}, the RH problem \eref{RHPNt}
can be solved similarly to the RH problem \eref{RHPsl}. Introduce
the following notation for the analytic branches of $N(k)$:
\begin{equation}
N(k)=\left\{
\begin{array}{ll}
N_R(k)\qquad & k\in\mathcal D_R\\
N_L(k) \qquad & k \in \mathcal D_L\\
N_{\infty}(k)\qquad &k\in\mathbb{C}\setminus(\mathcal
D_R\cup\mathcal D_L)
\end{array}
\right.. \label{Nbranch}
\end{equation}
Using the same arguments as in deriving equation
\eref{scgeneral} from \eref{jrR} one gets
\begin{equation}
N_\infty(k)=I+\frac{A_R}{k-1}+\frac{A_L}{k+1} \qquad k\in\mathbb C
\label{NA}
\end{equation}
where the matrices $A_R$ and $A_L$ do not depend on $k.$ Dropping
terms of order $x^{-1-\epsilon}$ and higher, we find
\begin{equation}
\fl
A_R=
\frac{2}{\kappa}
\left(
\begin{array}{cc}
\varkappa_L^2 -\frac\rmi{2x}\left(\frac\eta{2\pi}\right)^2
\frac{k_s}{k_s-1} & -\rmi\varkappa_R \rme^{\rmi\frac\eta2}
\sin\frac\eta2 \rme^{-\rmi t}
\left(\frac{k_s+1}{k_s-1}\right)^{\frac{\pi+\eta}{2\pi}}
\\
-\rmi\varkappa_L \rme^{-\rmi \frac \eta 2} \csc\frac\eta2
\rme^{\rmi t} \left(\frac{k_s+1}{k_s-1}
\right)^{\frac{\pi - \eta}{2 \pi }} &
\varkappa_R^2+\frac\rmi{2x}\left(\frac\eta{2\pi}\right)^2
\frac{k_s}{k_s-1}
\end{array}
\right) \label{ARt}
\end{equation}
and
\begin{equation}
\fl
A_L=
\frac{2}{\kappa}
\left(
\begin{array}{cc}
-\varkappa_R^2 -\frac\rmi{2x}\left(\frac\eta{2\pi}\right)^2
\frac{k_s}{k_s+1} &
\rmi\varkappa_L\rme^{\rmi\frac\eta2}\sin\frac\eta2
\rme^{-\rmi t}
\left(\frac{k_s+1}{k_s-1}\right)^{-\frac{\pi-\eta}{2\pi}}
\\
\rmi\varkappa_R \rme^{-\rmi \frac \eta 2} \csc \frac\eta 2
\rme^{\rmi t} \left(\frac{k_s+1}{k_s-1}
\right)^{-\frac{\pi+\eta}{2\pi}} & -\varkappa_L^2+\frac\rmi{2x}
\left(\frac{\eta}{2 \pi}\right)^2 \frac{k_s}{k_s+1}
\end{array}
\right). \label{ALt}
\end{equation}
Here
\begin{equation}
\fl \varkappa_R= \frac{\pi \rme^{\rmi x} (2
x)^{-1-\frac{\eta}{\pi}}} {\Gamma^2\left(-\frac{\eta}{2
\pi}\right) \sin \frac{\eta}{2}} \left( \frac{k_s^2}{k_s^2-1}
\right)^\frac{\pi +\eta}{2 \pi } \qquad \varkappa_L = \frac{\pi
\rme^{-\rmi x} (2 x)^{-1+\frac{\eta}{\pi}}}
{\Gamma^2\left(\frac{\eta}{2 \pi}\right) \sin \frac{\eta}{2}}
\left( \frac{k_s^2}{k_s^2-1} \right)^\frac{\pi -\eta}{2 \pi }
\label{kappat}
\end{equation}
and $\kappa$ is defined by equation \eref{Delta}. The matrices $N_R(k)$ and
$N_L(k)$ are found from (\ref{RHPNt}.b)
\begin{equation}
N_R(k)=N_{\infty}(k)[\tilde\theta_R(k)]^{-1}\qquad
N_L(k)=N_{\infty}(k)[\tilde\theta_L(k)]^{-1}. \label{NRL}
\end{equation}

Next, solve the RH problem \eref{RHPMt}. Use the same method as
was exploited in solving the RH problems \eref{RHPNt} and \eref{RHPsl}.
Denote the analytic branches of $M(k)$ by
\begin{equation}
M(k)=\left\{\begin{array}{ll}
M_s(k)\qquad &k\in \mathcal D_s \\
M_{\infty}(k)\qquad &k\in \mathbb{C}\setminus \mathcal D_s
\end{array}\right.. \label{Mbranch}
\end{equation}
The jump matrix $N_\infty\tilde\theta_s N_\infty^{-1}$ in
(\ref{RHPMt}.b) has a simple pole at $k=k_s.$ Therefore
\begin{equation}
M_{\infty}(k)=I+ \frac{A_s}{k-k_s} \qquad k\in\mathbb C
\label{MA}
\end{equation}
where the matrix $A_s$ does not depend on $k.$ Choosing
\begin{equation}
A_s=N_{\infty}(k_s)\upsilon_s(k_s) [N_\infty(k_s)]^{-1}
\label{Ast}
\end{equation}
one solves \eref{RHPMt}. This can be checked directly.

\subsection{Riemann-Hilbert problem \eref{RHPtime}: results in the large $x$ limit \label{approxt}}
Like the $t=0$ case, the matrix $K$ in the decomposition
\eref{U=KSQt} does not contribute to the asymptotic solution of
the RH problem \eref{RHPtU} to the order we are interested in.
Substituting \eref{StStilde} into the
decomposition \eref{U=KSQt} one gets the estimate
\begin{equation}
U=\tilde SQ[I+\mathcal O(x^{-1-\epsilon})] \label{schit}
\end{equation}
uniform outside an arbitrarily small vicinity of the contours
$\partial\mathcal D_{R,L,s}$ and $\Sigma_{1,3,s}\cap\mathcal
D_\infty.$ Using \eref{Z=UL} and \eref{Y=ZV} one gets the
approximation to $Y(k)$ up to the order of $x^{-1-\epsilon},$
uniform outside this vicinity. This completes the solution to the
RH problem \eref{RHPtime} in the large $x$ limit.


\section{Time dependent correlation functions: space-like region \label{spacelikeG}}
In this section we derive the asymptotic expressions for the
correlation functions $G_h(x,t)$ and $G_e(x,t)$ in the space-like
region $|x|>2|t|.$ The asymptotic expressions are derived assuming
that $x,t \to \infty$ at a fixed value of the parameter $k_s$
defined in \eref{ksdef}. The explicit expressions are written
assuming $x>0$ and $t>0;$ all other cases follow from the
relations \eref{xtominusx} and \eref{ttominust}.

All the calculations will be similar to those for the $t=0$ case.
In subsection \ref{Basymptt} we calculate the functions $B_{ab}$
and $C_{ab}$ \eref{BabCab} from the large $k$ expansion
\eref{RHPexpansion} of $Y(k).$ Having these functions we calculate
the Fredholm determinant $\det(\hat I+\hat V)$ in the large $x$
limit, see equation \eref{logdett}. The constant $C(\eta)$
entering \eref{logdett} is calculated in subsection \ref{casyt}.
This constant is found to be the same as in the $t=0$ case. We
give the answer for the correlation functions $G_h(x,t)$ and
$G_e(x,t)$ in subsection \ref{answerx}.

\subsection{Large $x$ asymptotics of $B_{ab},$ $C_{ab}$
and of the Fredholm determinant $\det(\hat I+\hat V)$
\label{Basymptt}}

In this subsection we use the results of section
\ref{section:RHPtime} on the asymptotic solution of the RH problem
\eref{RHPtime} to calculate the functions $B_{ab},$ $C_{ab}$ and
the Fredholm determinant $\det(\hat I+\hat V)$ in the large $x$
limit.

 It follows from \eref{schit},
\eref{Y=ZV}, \eref{Z=UL}, \eref{Qdeft}, \eref{S=MN},
\eref{Nbranch} and \eref{Mbranch} that in the vicinity of
$k=\infty$ the following uniform approximation is valid:
\begin{equation}
Y=M_{\infty} N_\infty Q_\infty V+{\cal O}(x^{-1-\epsilon}).
\label{MNQVinftyt}
\end{equation}
Recall that the matrix $M_\infty$ is given by equation \eref{MA},
the matrix $N_\infty$ by equation \eref{NA}, the matrix $Q_\infty$
by \eref{Ndef}, the matrix $V$ by \eref{V}. The approximation
\eref{MNQVinftyt} being uniform, the functions $B_{ab}$ and
$C_{ab}$ \eref{BabCab} can be calculated in the large $x$ limit
from the large $k$ expansion \eref{RHPexpansion} of equation
\eref{MNQVinftyt}. In particular,
\begin{equation}
\fl B_{--}= 2 \rmi\rme^{\rmi t} \rme^{-\frac{\rmi\eta}2}
\csc\frac\eta2 \left[
\left(\frac{k_s-1}{k_s+1}\right)^{-\frac{\pi-\eta}{2\pi}}\varkappa_L-
\left(\frac{k_s-1}{k_s+1}\right)^{\frac{\pi+\eta}{2\pi}}\varkappa_R
\right]\left[1+\mathcal O(x^{-\epsilon})\right]. \label{Bmmt}
\end{equation}
The function $b_{++},$ entering \eref{Gelectronintz} has the form
\begin{eqnarray}
\fl b_{++}=\left\{2 \rmi \rme^{-\rmi
t}\rme^{\frac{\rmi\eta}2}\sin\frac\eta2 \left[
\left(\frac{k_s-1}{k_s+1}\right)^{\frac{\pi-\eta}{2\pi}}\varkappa_L
-\left(\frac{k_s-1}{k_s+1}\right)^{-\frac{\pi+\eta}{2\pi}}\varkappa_R
\right]\right. \nonumber\\
\lo-\left.2\sin^2\frac\eta2\left(\frac{k_s-1}{k_s+1}\right)^{\frac\eta\pi}G_0
\right\}\left[ 1+\mathcal O(x^{-\epsilon})\right].
\end{eqnarray}
The expressions for the other potentials are quite bulky, so we
write down the differential equations \eref{xderiv} and
\eref{tderiv} in the large $x$ limit omitting intermediate
calculations
\begin{eqnarray}
\fl
\partial_x \ln \det(\hat I+\hat V)=
\frac{\rmi \eta}{\pi}- \frac{ k_s^2}{k_s^2-1}\frac{\eta^2}{2 \pi^2 x}
+2\rmi\kappa^{-1}(\varkappa_R^2-\varkappa_L^2)
\nonumber
\\
\lo- \frac{(k_s-1)\kappa+2}{(k_s^2-1)\kappa}\alpha_R
-\frac{(k_s+1)\kappa-2}{(k_s^2-1)\kappa}\alpha_L+ \mathcal
O\left(x^{-1-\epsilon}\right) \label{dxdet}
\end{eqnarray}
and
\begin{equation}
\partial_t \ln \det(\hat I +\hat V)=
\frac{2 k_s}{k_s^2-1} \frac{\eta^2}{2 \pi^2 x }+ \alpha_R
+\alpha_L+\mathcal O\left(x^{-1-\epsilon} \right) \label{dtdet}
\end{equation}
where
\begin{eqnarray}
\alpha_R=2 \rmi \rme^{\rmi t} (1-\rme^{-\rmi \eta}) \varkappa_R
G_0 \kappa^{-1}
\left(\frac{k_s-1}{k_s+1}\right)^\frac{3(\pi+\eta)}{2\pi}
\\
\alpha_L=-2 \rmi\rme^{\rmi t} (1-\rme^{-\rmi\eta}) \varkappa_L
G_0\kappa^{-1} \left(\frac{k_s-1}{k_s+1}\right)^{-\frac{3(\pi
-\eta)}{2\pi}}
\end{eqnarray}
and $\kappa$ is given by equation \eref{Delta}.

Integrating equations \eref{dxdet} and \eref{dtdet} asymptotically
one gets
\begin{equation}
\ln \det (\hat I +\hat V)= \frac{\rmi \eta}{\pi} x
-\left(\frac\eta{2\pi}\right)^2 \ln\left[x^2
\left(1-k_s^{-2}\right)\right] + C(\eta) + \mathcal
O\left(x^{-\epsilon}\right). \label{logdett}
\end{equation}
Note that $x^2(1-k_s^{-2})=x_Rx_L.$ We calculate the integration
constant $C(\eta)$ in subsection \ref{casyt}.
It will be the same as for the $t=0$ case.

\subsection{Calculation of $C(\eta)$ \label{casyt}}
In this subsection we calculate the constant $C(\eta)$ in
\eref{logdett} using the differential equation
\eref{diffetalndet}. The calculation will resemble that of
subsection \ref{casy}.

Consider the first term in the right hand side of equation
\eref{diffetalndet}. Split the integral in equation
\eref{detavect} in two parts and define
\begin{eqnarray}
J_R=\frac{1}{1-\rme^{-\rmi \eta}}\int_{0}^{1} \rmd k [\vec
f(k)]^{T} \sigma_2\partial_k \vec f(k)
\label{JRdef} \\
J_L=\frac{1}{1-\rme^{-\rmi \eta}}\int_{-1}^{0} \rmd k [\vec
f(k)]^T \sigma_2\partial_k \vec f(k). \label{JLdef}
\end{eqnarray}
Consider $J_R$ first. The large $x$ asymptotics of the function
$\vec f(k)$ in equation \eref{JRdef} is calculated as follows. Use
equation \eref{f=Ye}
\begin{equation}
\vec f(k)=Y_{+}(k)\vec e(k) \qquad k\in\Sigma_2. \label{f=Yex}
\end{equation}
Substituting \eref{Y=ZV} and \eref{V} into \eref{f=Yex} and using
\eref{edefs} and \eref{Epmprop} one gets
\begin{equation}
\vec f(k)= Z_{+}(k) \vec e_0(k) \qquad k\in\Sigma_2 \label{f=Ze0}
\end{equation}
where
\begin{equation}
\vec e_0(k) =\left(
\begin{array}{c}
e_{+}^0(k) \\
e_{-}(k)
\end{array} \right)
\qquad e_{+}^0(k) = \frac{\rmi(1-\rme^{\rmi\eta})}{2\sqrt\pi}
\rme^{-\tau(k)/2} \label{e0def}
\end{equation}
the matrix $Z(k)$ solves the RH problem \eref{RHPtZ} and the
functions $e_{-}(k)$ and $\tau(k)$ are defined by equation
\eref{edefs}. It follows from \eref{Z=UL} and \eref{Ldeft} that
\begin{equation}
Z_{+}(k)=U_{+}(k) \mu_1 (k) \qquad k \in \Sigma_2 \label{Z=Umu}
\end{equation}
where $U(k)$ solves the RH problem \eref{RHPtU}. Substituting
equation \eref{Z=Umu} into \eref{f=Yex} and performing the
analytic continuation of the formula \eref{schit} in complete
analogy with the analytic continuation of the formula \eref{schi}
discussed in subsection \ref{casy} and shown in figure
\ref{Fig:OmegaRcontinuation} we obtain the approximation to $\vec
f(k)$
\begin{equation}
\vec f=M_{\infty}N_R\vec g_R[I + \mathcal O(x^{-1-\epsilon})]
\label{fas}
\end{equation}
uniform for $k\in[0,1].$ The matrices $M_\infty$ and $N_R$ are
defined in subsection \ref{Sast}, the function $\vec g$ is given
by the formula
\begin{equation}
\vec g_R(k)=\left(
\begin{array}{c}
g^R_+(k) \\
g^R_-(k)
\end{array} \right)=
[Q_R(k)]_{+}\mu_1(k)\vec e_0(k)\qquad k\in [0,1]. \label{gRdef}
\end{equation}
Substituting into \eref{gRdef} the expression for $Q_R(k),$
defined by equations \eref{QRt}, \eref{TRLdef} and
\eref{phientries}, and using the identity \eref{F11iden} one gets
\begin{equation}
\fl \eqalign{ g_{+}^R(k;x,t)= -
\frac{\sqrt\pi\rme^{\frac{\rmi\eta}4}\rme^{{\tau(k)}/2-\tau(1)}}
{\Gamma\left(-\frac\eta{2\pi}\right)}
\frac{{}_1F_1\left[\frac\eta{2\pi}+1, 1; \rmi x_R (\lambda_R-1)
\right]} {\left[x_R (k+1)\right]^{\frac{\eta} {2 \pi}}}
\left(\frac{k-1}{\lambda_R-1} \right)^\frac{\eta}{2 \pi}
\\
g_{-}^R(k;x,t)= \frac{\sqrt{\pi}\rme^{-\frac{\rmi \eta}{4}}
\rme^{{\tau(k)}/{2}}}
{\Gamma\left(\frac{\eta}{2\pi}\right)\sin\frac{\eta}{2} } \frac{
{}_1F_1\left[\frac{\eta}{2\pi}, 1; \rmi x_R (\lambda_R-1)
\right]}{\left[ x_R (k+1)\right]^{-\frac \eta {2 \pi }}}
\left(\frac{k-1}{\lambda_R-1} \right)^{-\frac{\eta}{2 \pi}} }
\label{gexplicitt}
\end{equation}
where $\lambda_R=\lambda_R(k)$ is defined by equation \eref{zdef}.

Using the uniform estimate \eref{fas} and equations
\eref{gexplicitt}, \eref{NA}-\eref{Ast} one can derive the
following estimate for \eref{JRdef}
\begin{equation}
J_R=\frac{1}{1-\rme^{-\rmi\eta}}\int_0^1 \rmd k [\vec g_R(k)]^T
\sigma_2 \partial_k \vec g_R(k)[1+ \mathcal O (x^{-1-\epsilon})].
\label{JRas}
\end{equation}
Performing the integral in \eref{JRas} asymptotically one gets
\begin{equation}
J_R= \frac{\rmi x_R}{2 \pi}- \frac{\eta}{2\pi^2} \ln x_R+
\frac12C_1(\eta)+\mathcal O(x^{-\epsilon}) \label{JRanswer}
\end{equation}
where $C_1(\eta)$ is defined by equation \eref{C1}.

The large $x$ asymptotics of $J_L$ is calculated similarly,
yielding
\begin{equation}
J_L= \frac{\rmi x_L}{2 \pi}- \frac{\eta}{2\pi^2} \ln x_L+
\frac12C_1(\eta)+\mathcal O(x^{-\epsilon}).
\end{equation}

We have calculated the large $x$ asymptotics of the first term in
the right hand side of equation \eref{diffetalndet}. By analysis
of equation \eref{W} one can show that the second term in the
right hand side of equation \eref{diffetalndet} is of the order of
$x^{-\epsilon}.$ Therefore, the large $x$ asymptotics of
\eref{diffetalndet} is
\begin{eqnarray}
\fl\partial_\eta \ln \det (\hat I +\hat V) &=J_R+J_L + \mathcal O
(x^{-\epsilon})\nonumber\\
&=\frac{\rmi x}\pi-\frac\eta{2\pi^2}\ln(x_Rx_L)+C_{1}(\eta)+
\mathcal O (x^{-\epsilon}). \label{estimateforconst}
\end{eqnarray}

Integrating equation \eref{estimateforconst} over $\eta,$ taking
into account the initial condition \eref{inicondeta} and comparing
the resulting expression with \eref{logdett} one finds that
$C(\eta)$ is given by the same expression \eref{Cdef} as in the
$t=0$ case.

\subsection{Time dependent correlation functions
in the space-like region: the results \label{answerx}}

In this subsection we write down the asymptotic expressions for
the correlation functions $G_h(x,t)$ and $G_e(x,t)$ in the
space-like region $|x|>2|t|.$ The asymptotic expressions are given
assuming that $x,t \to \infty$ at a fixed value of the parameter
$k_s$ defined in \eref{ksdef}. The explicit expressions are
written assuming $x>0$ and $t>0;$ all other cases follow from the
relations \eref{xtominusx} and \eref{ttominust}.

The correlation function $G_h(x,t)$ is given by the formula
\eref{Gholeintzdeformed}. The asymptotics of $B_{--}$ is given by
\eref{Bmmt}; the asymptotics of $\det(\hat I+\hat V)$ by
\eref{logdett} with the constant $C(\eta)$ given by \eref{Cdef}.
Recall that $z=\exp(\rmi \eta),$ therefore the point $z=1/2$
corresponds to $\eta=\rmi \ln 2.$ The leading contribution to the
asymptotics of $G_h(x,t)$ comes from the first term in the right
hand side of equation \eref{Gholeintzdeformed}, yielding
\begin{eqnarray}
\fl G_h(x,t)=\frac{\Xi}{2\rmi}\frac{\rme^{-x\ln2/\pi}}
{(x+2t)^{2\Delta}(x-2t)^{2\bar\Delta}}
\exp\left[\rmi(x-x_0)-\frac{\rmi\ln2}{\pi}\ln(x+2t) \right]
\nonumber
\\
\lo- \frac{\Xi}{2\rmi}\frac{\rme^{-x\ln2/\pi}}
{(x-2t)^{2\Delta}(x+2t)^{2\bar\Delta}}
\exp\left[-\rmi(x-x_0)+\frac{\rmi\ln2}{\pi}\ln(x-2t) \right]
\label{Gholespacelike}
\end{eqnarray}
where the normalization constant $\Xi$ is defined by equation
\eref{Xidef} and the phase shift $x_0$ by equation \eref{x0def}.
The anomalous exponents are given by
\begin{equation}
\Delta= \frac{1}2- \frac{1}8 \left(\frac{\ln2}\pi\right)^2 \qquad
\bar\Delta=- \frac{1}{8} \left(\frac{\ln 2}{\pi}\right)^2.
\label{deltasdef}
\end{equation}
The relative correction to \eref{Gholespacelike} is of the order
of $x^{-1}.$

Show that the second term in the right hand side of equation
\eref{Gholeintzdeformed} does not contribute to the asymptotics
\eref{Gholespacelike}. By virtue of \eref{logdett}, the
determinant $\det(\hat I+\hat V)$ is of the order of $\exp(x\ln
r/\pi)$ on the integration contour $\gamma,$ where $\gamma$ is the
circle $|z|=r,$ figure~\ref{Fig:zintegral}. Therefore, for $r<1/2$
and for sufficiently large $x$ the contribution of the second term
is negligible.

The correlation function $G_e(x,t)$ is calculated similarly,
yielding
\begin{equation}
\fl G_e(x,t)=-G_h^*(x,t)+\rme^{\rmi
t}G_0(x,t)\frac{\rme^{-x\ln2/\pi}\rme^{C(\rmi\ln2)}}
{(x^2-4t^2)^{2\bar\Delta}}
\left(\frac{x-2t}{x+2t}\right)^{\rmi\ln2/\pi}
\label{Gelectronspacelike}
\end{equation}
where $G_h(x,t)$ is given by equation \eref{Gholespacelike}, the
function $G_0(x,t)$ is defined by equation \eref{Gvac} and $\bar
\Delta$ defined by \eref{deltasdef}.

Compare the result \eref{Gelectronspacelike} with the general
relation \eref{GeGhanticom}. One can see that the last term in the
right hand side of equation \eref{Gelectronspacelike} corresponds
to the averaged anticommutator of fermion fields in equation
\eref{GeGhanticom}. It contains a rapidly oscillating factor
$\exp(\rmi t).$ This means that although formally present in the
long distance expansion of the correlation function $G_e$ this
term is irrelevant at energies smaller than the chemical
potential.

\section{Riemann-Hilbert problem at $t\neq0.$ Time-like region.}

In this section we find the asymptotic solution of the
Riemann-Hilbert problem \eref{RHPtime} for $|x|<2|t|.$ We assume
that $x>0$ and $t>0.$ We will consider the large $t$ asymptotics
at a fixed value of the parameter $k_s$ defined in \eref{ksdef}.

In subsection \ref{reformulationtl} we reformulate the RH problem
\eref{RHPtime} so as to make it similar to the RH problem \eref{RHPx}.
In subsection 7.2 we repeat the
construction made for the $t=0$ case in subsections
\ref{deformed1} and \ref{deformed2}. In
subsection 7.3 we give an explicit solution to the jump relation
(7.7.b) in a vicinity of the points $k=\pm1$ and $k=k_s.$ In
subsection 7.4 we perform the matching procedure analogous to the one
of subsection \ref{perturthe}.

\subsection{Reformulation of the Riemann-Hilbert problem \eref{RHPtime} \label{reformulationtl} }
In this subsection we reformulate the RH problem \eref{RHPtime} so
as to make use of the results of subsections \ref{lenstsl2} and
\ref{QRLt}.

\begin{figure}
\centering
\includegraphics[width=0.4\textwidth]{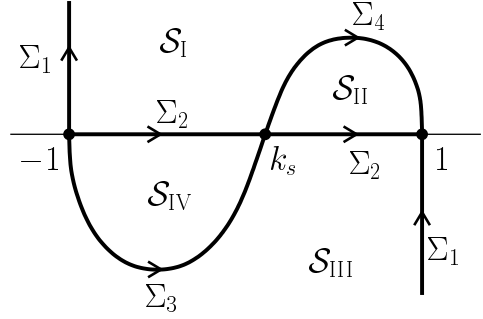}
\caption{The conjugation contour
$\Sigma_1\cup\Sigma_2\cup\Sigma_3\cup\Sigma_4$ for the RH problem
\eref{RHPtlZ}. The contour $\Sigma_1$ consists of two
semi-infinite lines parallel to the imaginary axis.}
\label{Fig:tlikecontour}
\end{figure}

Consider a contour shown in figure \ref{Fig:tlikecontour}.
This contour divides the complex $k$-plane into four regions,
denoted as $\mathcal S_{\rm I}$ through $\mathcal S_{\rm IV}.$
Represent the solution $Y(k)$ of the RH problem \eref{RHPtime} as
a product of two matrices
\begin{equation}
Y(k)=Z(k)V(k) \label{tlY=ZV}
\end{equation}
where
\begin{equation}
\fl V(k)= \left( \begin{array}{cc} 1&0\\ 2 \rmi
\rme^{\tau(k)}\varrho_1(k)&1 \end{array} \right) \left(
\begin{array}{cc}
1&-2 \sin^2 \frac{\eta}{2} [E(k)-\rmi \rme^{-\tau(k)}\varrho_2(k)]
\\ 0&1 \end{array} \right). \label{tlVdef}
\end{equation}
The piecewise constant functions $\varrho_1$ and $\varrho_2$ are
defined by
\begin{equation} \varrho_1(k)=\left\{ \begin{array}{cl}
\rme^{\rmi \eta}&\qquad k\in \mathcal S_{\rm II} \\ -\rme^{-\rmi
\eta}&\qquad k \in \mathcal S_{\rm IV}\\ 0&\qquad {\rm otherwise}
\end{array} \right. \label{tlsigmadef} \end{equation}
and
\begin{equation}
\varrho_2(k)=\left\{ \begin{array}{cll} 1&\qquad k\in \mathcal
S_{\rm III} &\qquad {\rm Im}k>0 \\ -1&\qquad k \in \mathcal S_{\rm
I} &\qquad {\rm Im}k<0\\ 0&\qquad {\rm otherwise}& \end{array}
\right. . \label{tlrhodef}
\end{equation}
The contour $\Sigma_1$ is chosen so that the real part of
$\tau(k)$ is positive for all $k\in \Sigma_1$ except $k=\pm 1$ and
\begin{equation}
\mathrm{Re}\tau(k)\to +\infty \qquad {\rm as } \qquad k\to \infty
\qquad k\in \Sigma_1 \label{tauto01}
\end{equation}
Note that the condition \eref{tauto01} implies
\begin{equation}
V(k)\to I \qquad{\rm as } \qquad  k\to \infty. \label{Vastl}
\end{equation}

Using equations \eref{RHPtime}, \eref{tlY=ZV} and \eref{tlVdef},
and the property \eref{Vastl} one gets the following RH problem
for $Z(k)$
\begin{eqnarray}
\eqalign{\mbox{(a)}\quad& Z(k) \mbox{ is  analytic  in } {\mathbb
C} \setminus
\Sigma \\
\mbox{(b)}\quad& Z_{+}(k)=Z_{-}(k)\mu_{i}(k) \qquad  k \in
\Sigma_{i} \qquad
i=1,2,3,4 \\
\mbox{(c)}\quad& Z(k)\to I \mbox{ as } k\to \infty} \label{RHPtlZ}
\end{eqnarray}
where the jump matrices $\mu_1,$ $\mu_2$ and $\mu_3$
are given by equations \eref{mu1def}, \eref{mu2def} and
\eref{mu3def}, respectively. The matrix $\mu_4$ is defined as
\begin{equation}
\mu_4=\left( \begin{array}{cc} 1&0 \\
2\rmi \rme^{\tau(k)}\rme^{\rmi \eta}& 1 \end{array} \right).
\label{mu4def}
\end{equation}

\subsection{Factorization of the Riemann-Hilbert problem \eref{RHPtlZ}}
In this subsection we sketch the scheme of the analysis of the RH
problem \eref{RHPtlZ} in the large $t$ limit. This scheme is
similar to the one given in subsection \ref{lenstsl2}.

\begin{figure}
\centering \includegraphics[width=0.5\textwidth]{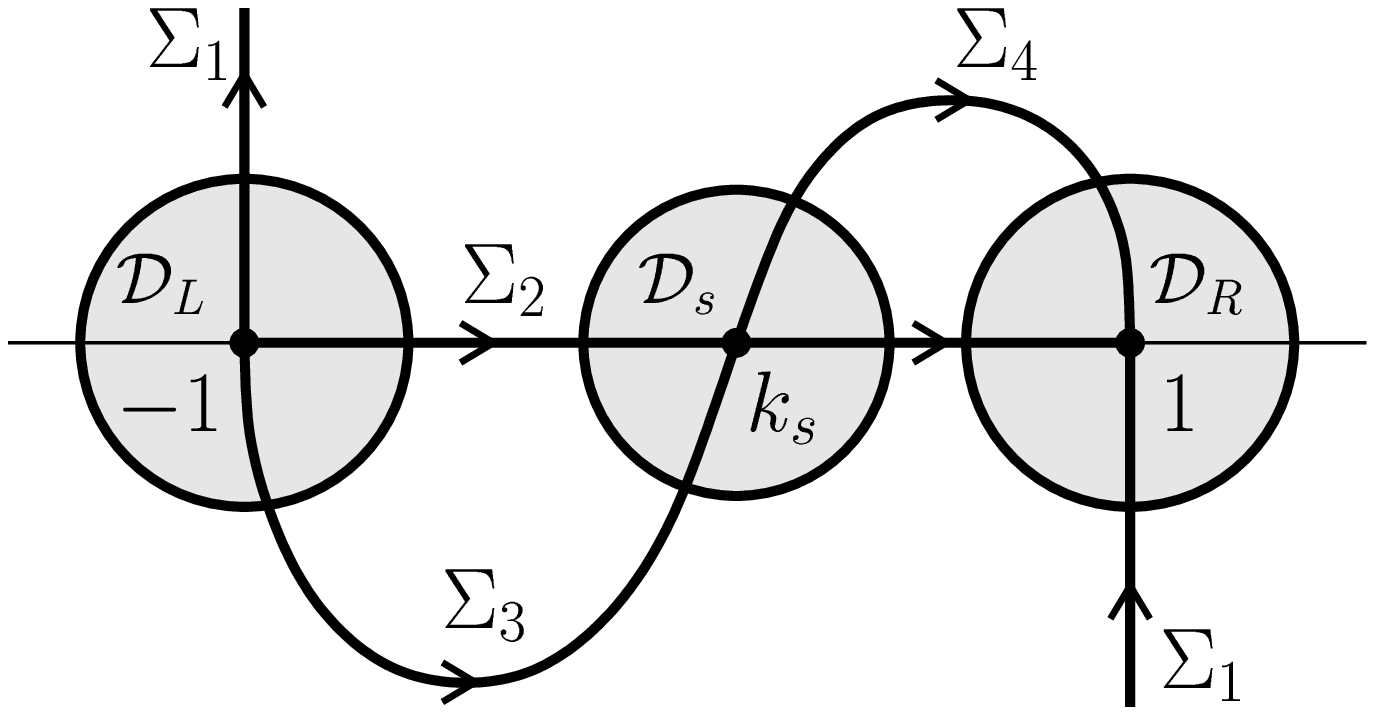}
\caption{The contour used in the factorization of the RH problem
\eref{RHPtlZ}. The discs $\mathcal D_R,$ $\mathcal D_L$ and
$\mathcal D_s$ are shaded in gray.} \label{Fig:tlksplit}
\end{figure}

The matrices $\mu_1,$ $\mu_3$ and $\mu_4$ converge to the identity
matrix, as $t\to\infty,$ everywhere on their conjugation contours
except at the points $k=\pm1$ and $k=k_s.$ Introduce three disks
$\mathcal D_L,$ $\mathcal D_R$ and $\mathcal D_s$ as shown in
figure \ref{Fig:tlksplit}.
Denote the domain $\mathbb C \setminus(\mathcal D_L \cup \mathcal
D_R\cup \mathcal D_s)$ as $D_\infty.$ Represent the solution
$Z(k)$ of the RH problem \eref{RHPtlZ} as a product of three
matrices
\begin{equation}
Z=KSQ. \label{Z=KSQ}
\end{equation}
Here the matrix $Q(k)$ is defined by
\begin{equation}
Q(k)=Q_\alpha(k) \qquad k\in \mathcal D_\alpha \qquad \alpha =L,
R, s, \infty. \label{Qproblemtl}
\end{equation}
The matrix $Q_{\infty}$ is given by equation \eref{Qdef}. The
matrices $Q_R$ and $Q_L$ are defined in $\mathcal D_R$ and
$\mathcal D_L,$ respectively, and satisfy there the jump relations
\begin{equation}
\left[Q_R(k)\right]_{+}=\left[Q_R(k)\right]_{-} \mu_i(k) \qquad
k\in \Sigma_i\cap \mathcal D_R \qquad i = 1, 2, 4 \label{QRttl}
\end{equation}
and
\begin{equation}
\left[Q_L(k)\right]_{+}=\left[Q_L(k)\right]_{-} \mu_i(k) \qquad
k\in \Sigma_i\cap \mathcal D_L \qquad  i= 1, 2, 3 \label{QLttl}.
\end{equation}
The matrix $Q_s(k)$ is defined in $\mathcal D_s$ and satisfies
\begin{equation}
\left[Q_s(k)\right]_{+}=\left[Q_s(k)\right]_{-} \mu_i(k) \qquad
k\in \Sigma_i\cap \mathcal D_s \qquad i = 2, 3, 4.
\label{Qalphatl}
\end{equation}
We give solutions to the jump relations \eref{QRttl}, \eref{QLttl}
and \eref{Qalphatl} in subsection \ref{exacttl}.

The matrix $S(k)$ in the decomposition \eref{Z=KSQ} is the
solution to the RH problem \eref{RHPst} with the conjugation
matrices $\theta_{R,L,s}$ defined by equation \eref{thetaRLsdef}.
The matrices $Q_R,$ $Q_L$ and $Q_s$ in \eref{thetaRLsdef} solve
the jump relations \eref{QRttl}, \eref{QLttl} and \eref{Qalphatl},
respectively.

Like the spacelike case, the matrix $K$ can be replaced by the
identity matrix.
\subsection{Exact solutions to the jump relations
\eref{QRttl}, \eref{QLttl} and \eref{Qalphatl} \label{exacttl}}

In this subsection we solve explicitly the jump relations
\eref{QRttl}, \eref{QLttl} and \eref{Qalphatl}. We use the results
of subsection \ref{QRLt}.

The jump relation \eref{QLttl} coincides with the jump relation
\eref{QjumptL}. Therefore one can use the solution \eref{QLt} to
this relation. To obtain a solution to the jump relation
\eref{QRttl} consider the conjugation matrices $\mu_1,$ $\mu_2$
and $\mu_4$ defined on the contours $\Sigma_1,$ $\Sigma_2$ and
$\Sigma_4,$ shown in figure \ref{Fig:tlksplit}. These matrices
satisfy the following involutions
\begin{equation}
\begin{array}{rl}
\mu_1(k;x,t, \eta)&=P_1^{-1}\mu_3(k; -x, -t, -\eta)P_1 \\ \mu_2(k;
\eta)&=P_1^{-1}\mu_2(k; -\eta)P_1 \\ \mu_4(k; x,t,
\eta)&=P_1^{-1}\mu_1(k; -x, -t, -\eta)P_1 \end{array}
\label{involutiontl} \end{equation} where $P_1$ is a
$k$-independent matrix defined by \begin{equation} P_1=\left(
\begin{array}{cc} 0& -\rmi \rme^{-\rmi \eta/2}\sin\frac{\eta}{2}\\
\rmi \rme^{\rmi \eta/2}\csc\frac{\eta}{2}&0 \end{array} \right).
\label{P1} \end{equation} Denote the solution \eref{QRt} to the
jump relation \eref{QjumptR} by $Q_R^{\rm sl} (k;x,t, \eta).$
Comparing the conjugation contours in the disks $\mathcal D_R$
shown in figures \ref{Fig:tsplit} and \ref{Fig:tlksplit} and using
the involutions \eref{involutiontl} a solution $Q_R$ to the jump
relation \eref{QRttl} can be found from $Q_R^{\rm sl}$
\begin{equation}
Q_R(k;x,t,\eta)=P_1 Q_R^{\rm sl} (k;-x,-t, -\eta) P_1^{-1}
\label{QRtl} \end{equation}

Finally, write down a solution to the jump relation
\eref{Qalphatl}:
\begin{equation}
Q_s(k)= \left(
\begin{array}{cc}
1&0\\
q(k)&1
\end{array}
\right) Q_\infty(k) \label{Qstl}
\end{equation}
where
\begin{equation}
q(k)=\int_{\Sigma_s} \frac{\rmd p}{2\pi \rmi}
\left(\frac{1-p}{1+p}\right)^{-\frac{\eta}{\pi}} \frac{2\rmi
\rme^{\tau(p)}}{p-k} \label{qdeftl}
\end{equation}
and the integral runs along the contour $\Sigma_s$ which consists
of the part of the contour $\Sigma_3$ lying outside $\mathcal D_L$
and the part of the contour $\Sigma_4$ lying outside $\mathcal
D_R,$ see figure \ref{Fig:tlksplit}.
\subsection{Solution to the Riemann-Hilbert problem \eref{RHPst} in the large $t$ limit}

In this subsection we solve the RH problem \eref{RHPst} in the
large $t$ limit. The procedure will resemble that of subsection
\ref{Sast}.

The conjugation matrix $\theta_{L}$ is approximated by a matrix
$\tilde \theta_L$
\begin{equation}
\theta_{L}=\tilde \theta_{L}+\mathcal O
\left(t^{-1-\epsilon}\right) \label{thetaLastl}
\end{equation}
uniformly for $k\in\partial \mathcal D_L,$ where $\tilde \theta_L
$ is defined by equations \eref{thetaLast} and \eref{upsilonLt}.
The conjugation matrix $\theta_R$ is approximated by a matrix
$\tilde \theta_R$
\begin{equation}
\theta_{R}=\tilde \theta_{R}+\mathcal O
\left(t^{-1-\epsilon}\right)
\end{equation}
uniformly for $k\in \partial \mathcal D_R,$ where the matrix $\tilde \theta_R$ is
given by equation \eref{thetaRast} and  $\upsilon_{R}$ is defined
as
\begin{eqnarray} \upsilon_R(k;x, t,
\eta)= -P_1 \upsilon_R^{\rm sl}(k;-x,-t, -\eta)P_1.
\label{upsilonRtl}
\end{eqnarray}
Here we use the notation $\upsilon_R^{\rm sl}(k;x,t, \eta)$ for
the matrix given by equation \eref{upsilonRt}. The matrix $P_1$ is
defined by equation \eref{P1}. The conjugation matrix $\theta_s$
is approximated by a matrix $\tilde\theta_s$
\begin{equation}
\theta_s=\tilde\theta_s[I+\mathcal O(t^{-3/2})]
\end{equation}
uniformly for $k\in\partial\mathcal D_s.$ The matrix
$\tilde\theta_s$ is given by equation \eref{tildetetaSsl} where
$\upsilon_s(k)$ is calculated using \eref{thetaRLsdef},
\eref{Qstl} and \eref{qdeftl}
\begin{equation}
\upsilon_s=\left(
\begin{array}{cc}
0&0\\
-2 \left( \frac{1-k_s}{1+k_s}\right)^{-\frac{\eta}{\pi}} G^*_0 &0
\end{array}\right).
\end{equation}

The asymptotic estimate for the matrix $S$ is given by equation
\eref{StStilde} where the matrix $\tilde S$ is the solution of the
RH problem \eref{RHPstildet} with the matrices $\tilde
\theta_{R,L,s}$ defined in the paragraph above. Like the spacelike
case, solve the RH problem \eref{RHPstildet}  using the
decomposition \eref{S=MN}. The matrix $N_\infty$ entering the
definition \eref{Nbranch} is given by equation \eref{NA} with
matrices $A_R$ and $A_L$ given by
\begin{equation}
\fl A_R=\frac{2}{\kappa} \left(
\begin{array}{cc} \varkappa_L^2+\frac{\rmi}{4t}\left(\frac{\eta}{2\pi}\right)^2
\frac{1}{1-k_s} & -\varkappa_R\sin\frac{\eta}{2}
\rme^{-\rmi t} \left(\frac{1+k_s}{1-k_s}
\right)^\frac{\pi+\eta}{2 \pi} \\ -\varkappa_L\csc\frac{\eta}{2}
\rme^{\rmi t} \left(\frac{1+k_s}{1-k_s}
\right)^\frac{\pi-\eta}{2 \pi} &
\varkappa_R^2-\frac{\rmi}{4t}\left(\frac{\eta}{2\pi}\right)^2
\frac{1}{1-k_s}
\end{array}
\right) + \mathcal O(t^{-1-\epsilon})
\label{ARtl}
\end{equation}
and
\begin{equation}
\fl A_L=\frac{2}{\kappa} \left(
\begin{array}{cc} -\varkappa_R^2-\frac{\rmi}{4t}\left(\frac{\eta}{2\pi}\right)^2
\frac{1}{1+k_s} & -\varkappa_L\sin\frac{\eta}{2}
\rme^{-\rmi t} \left(\frac{1-k_s}{1+k_s}
\right)^\frac{\pi-\eta}{2 \pi} \\ -\varkappa_R\csc\frac{\eta}{2}
\rme^{\rmi t} \left(\frac{1-k_s}{1+k_s}
\right)^\frac{\pi+\eta}{2 \pi} &
-\varkappa_L^2+\frac{\rmi}{4t}\left(\frac{\eta}{2\pi}\right)^2
\frac{1}{1+k_s}
\end{array}
\right)+ \mathcal O(t^{-1-\epsilon}).
\label{ALtl}
\end{equation}
Here
\begin{equation}
\eqalign{ \varkappa_R= -\frac{\rmi \pi \rme^{\rmi x}
(4t)^{-1-\frac{\eta}{\pi}}} {\Gamma^2\left(-\frac{\eta}{2
\pi}\right)  \rme^{\frac{\rmi \eta}{2}} \sin \frac{\eta}{2}}
\left( \frac{1}{1-k_s^2} \right)^\frac{\pi +\eta}{2 \pi }
\\
\varkappa_L =- \frac{\rmi \pi \rme^{-\rmi x}
(4t)^{-1+\frac{\eta}{\pi}}} {\Gamma^2\left(\frac{\eta}{2
\pi}\right) \rme^{-\frac{\rmi \eta}{2}} \sin \frac{\eta}{2}}
\left( \frac{1}{1-k_s^2} \right)^\frac{\pi -\eta}{2 \pi } }
\label{kappatl}
\end{equation}
and $\kappa$ is defined by equation \eref{Delta}. The matrix
$M_\infty(k)$ in the decomposition \eref{S=MN} is given by
equation \eref{MA} with $A_s$ defined by equation \eref{Ast}.

\section{Time dependent correlation function: time-like region, $k_s\neq 0$
\label{answertl}} In this section we derive the asymptotics of the
correlation functions $G_e(x,t)$ and $G_h(x,t)$ in the time-like
region $|x|<2|t|.$ The asymptotic expressions will be given
assuming that $x,t\to \infty$ at a fixed value of the parameter
$k_s$ defined in equation \eref{ksdef}. It will also be assumed
that $x>0$ and $t>0$; all other cases follow from the relations
\eref{xtominusx} and \eref{ttominust}.

The calculations in this section are similar to those given in
sections \ref{zerotimeG} and \ref{spacelikeG}. We will therefore
omit most technical details already explained in those sections
and focus on the particulars of the time-like case. In subsection
\ref{Basympttimelike} we give the the large $t$ asymptotics of
$B_{--},$ $b_{++}$ and $\det(\hat I +\hat V).$ In subsection
\ref{answert} we perform the contour integral entering
representations \eref{Gholeintz} and \eref{Gelectronintz} and
obtain the asymptotic expressions for the correlation functions
$G_e(x,t)$ and $G_h(x,t).$

\subsection{Large $t$ asymptotics of $B_{ab}$, $C_{ab}$ and of the Fredholm
determinant $ \det(\hat I+\hat V)$ \label{Basympttimelike}}

The asymptotic expressions for the functions $B_{--}$ and $b_{++}$
are calculated similarly to subsection
 \ref{Basymptt} yielding
\begin{eqnarray}
\fl B_{--} = \left\{ 2
\left(\frac{1+k_s}{1-k_s}\right)^\frac{\eta}{\pi} G_0^{*}
\right.
\nonumber\\
\lo+ \left.2\rme^{\rmi t}\csc\frac{\eta}2 \left[\left(
\frac{1-k_s}{1+k_s} \right)^{-\frac{\pi-\eta}{2\pi}} \varkappa_L +
\left( \frac{1-k_s}{1+k_s}  \right)^\frac{\pi+\eta}{2\pi}
\varkappa_R \right]\right\}\left[1+\mathcal
O(t^{-\epsilon})\right]
\label{Bmmtl}
\end{eqnarray}
and
\begin{equation}
\fl b_{++}= 2 \rme^{-\rmi t}\sin \frac{\eta}{2} \left[\left(
\frac{1-k_s}{1+k_s} \right)^\frac{\pi-\eta}{2\pi} \varkappa_L +
\left(\frac{1-k_s}{1+k_s}\right)^{-\frac{\pi+\eta}{2\pi}}
\varkappa_R \right]\left[1 + \mathcal O(t^{-\epsilon})\right]
\label{bpptl}
\end{equation}
where $\varkappa_R$ and $\varkappa_L$ are given by equaiton
\eref{kappatl}. The large $t$ asymptotics of the Fredholm
determinant $\det(\hat I +\hat V)$ is
\begin{equation}
\ln \det (\hat I +\hat V)= \frac{\rmi \eta}{\pi} x
-\left(\frac\eta{2\pi}\right)^2 \ln\left[4 t^2
\left(1-k_s^{2}\right)\right] + \tilde C(\eta) + \mathcal
O\left(t^{-\epsilon}\right). \label{logdett2}
\label{logdettl}
\end{equation}

The constant $\tilde C(\eta)$ can, in principle, be calculated in
the same manner as the constant $C(\eta)$ for the space-like
region. There is, however, a difference which did not allow us to
obtain a closed analytical expression for $\tilde C(\eta)$ so far.
Unlike the space-like case,  the second term in the right hand
side of equation \eref{diffetalndet} is not small as a function of
the asymptotic parameter $t$. Therefore, the calculation of
$\tilde C(\eta)$ requires the asymptotic analysis of the
expression \eref{W}, which is straightforward but cumbersome. This
is in contrast with the high temperature case, where the
calculation of such a constant was done using an additional
differential equation involving the functions $B_{ab}$ and
$C_{ab}$ only \cite{GIK-98}. Such an equation does not exist in
the zero temperature case, considered here.

\subsection{Time dependent correlation functions: time-like region, $k_s\neq0.$
The results \label{answert}}
In this subsection we write down the asymptotic expressions for
the correlation functions $G_h(x,t)$ and $G_e(x,t)$ in the
time-like region $|x|<2|t|.$ The asymptotic expressions are given
assuming that $x,t \to \infty$ at a fixed value of the parameter
$k_s$ defined in \eref{ksdef}. The explicit expressions are
written assuming $x>0$ and $t>0;$ all other cases follow from
the relations \eref{xtominusx} and \eref{ttominust}.

Consider the representations
\eref{Gholeintzdeformed} and \eref{Gelectronintzdeformed}.
Applying the same arguments as given in subsection \eref{answerx}
one can show that the contributions from the integrals along the
contour $\gamma$ in \eref{Gholeintzdeformed} and
\eref{Gelectronintzdeformed} can be neglected in the large $t$
limit. Substituting the results \eref{Bmmtl}, \eref{bpptl} and
\eref{logdettl} in equations \eref{Gholeintzdeformed} and
\eref{Gelectronintzdeformed} we find
\begin{eqnarray}
\fl G_h(x,t)=\frac{\tilde \Xi}{2\rmi}\frac{\rme^{-x\ln2/\pi}}
{(2t+x)^{2\Delta}(2t-x)^{2\bar\Delta}}
\exp\left[\rmi(x-x_0)-\frac{\rmi\ln2}{\pi}\ln(2t+x) \right]
\nonumber
\\
\lo+ \frac{\tilde \Xi}{4\rmi}\frac{\rme^{-x\ln2/\pi}}
{(2t-x)^{2\Delta}(2t+x)^{2\bar\Delta}}
\exp\left[-\rmi(x-x_0)+\frac{\rmi\ln2}{\pi}\ln(2t-x) \right]
\nonumber
\\
\lo+ \frac{\rme^{-\rmi t}}{2}G_0^*(x,t)
\frac{\rme^{-x\ln2/\pi}\rme^{C(\rmi\ln2)}}
{(4t^2-x^2)^{2\bar\Delta}}
\left(\frac{2t-x}{2t+x}\right)^{-\rmi\ln2/\pi}
\label{Gholetimelike}
\end{eqnarray}
and
\begin{eqnarray}
\fl G_e(x,t)=-\frac{\tilde \Xi}{\rmi}\frac{\rme^{-x\ln2/\pi}}
{(2t-x)^{2\Delta}(2t+x)^{2\bar\Delta}}
\exp\left[\rmi(x-x_0)-\frac{\rmi\ln2}{\pi}\ln(2t-x) \right]
\nonumber
\\
\lo- \frac{\tilde \Xi}{2\rmi}\frac{\rme^{-x\ln2/\pi}}
{(2t+x)^{2\Delta}(2t-x)^{2\bar\Delta}}
\exp\left[-\rmi(x-x_0)+\frac{\rmi\ln2}{\pi}\ln(2t+x) \right],
\label{Gelectrontimelike}
\end{eqnarray}
where the dimensions $\Delta$ and $\bar \Delta$ are given by equation \eref{deltasdef},
the phase shift $x_0$ by equation \eref{x0def}.  The normalization constant
$\tilde \Xi$ can be obtained from the constant $\Xi$ given in
equation \eref{Xidef} by replacing the constant $C(\eta)$ with $\tilde C(\eta).$

\section{Time-dependent correlation functions: time-like region, $x=0$ \label{ks=0}}
In this section we consider the special case of $x=0.$ The large
$t$ asymptotics of the correlation functions $G_e(0,t)$ and
$G_h(0,t)$ are given assuming that $t>0$; the case of negative $t$
follows from the relation \eref{ttominust}. In the derivation of
the asymptotical expressions we the use the results of section
\ref{answertl}.

For $x=0$ it is convenient to calculate the large $t$ asymptotics
of the integrals (2.6) and (2.7) without the deformation of the
integration contour leading to (2.32) and (2.33). Substituting
asymptotic formulae (8.2) and (8.3) in equation (2.7) we arrive at
the expression
\begin{eqnarray}
\fl G_e(0,t)=\int_{-\pi}^{\pi} \rmd \eta
\frac{F(\eta)}{1-\cos\eta} \frac{\rmi
4^{-1+\frac{\eta}{\pi}-\left(\frac\eta{2\pi}\right)^2}\rme^{\tilde
C(\eta)} } {\Gamma^2(\frac{\eta}{2 \pi}) \rme^{-\rmi \frac\eta2} }
\exp \left \{\left[-1+\frac{\eta}{\pi}-2
\left(\frac{\eta}{2\pi}\right)^2\right]\ln t \right \} \nonumber
\\
\lo+ \int_{-\pi}^{\pi} \rmd \eta \frac{F(\eta)}{1-\cos\eta}
\frac{\rmi
4^{-1-\frac{\eta}{\pi}-\left(\frac\eta{2\pi}\right)^2}\rme^{\tilde
C(\eta)} } {\Gamma^2(-\frac{\eta}{2 \pi}) \rme^{\rmi \frac\eta2} }
\exp \left \{\left[-1-\frac{\eta}{\pi}-2
\left(\frac{\eta}{2\pi}\right)^2\right]\ln t \right\}.
\label{electronsaddlepoit}
\end{eqnarray}
For large $t$ the integrals on the right hand side of (9.1) can be
evaluated in the saddle point approximation with $\ln t$ playing
the role of the asymptotic parameter. The relative correction to
the integrands in equation (9.1) is of the order of
$t^{-\epsilon}.$ At the saddle points, $\eta=\pm \pi,$ one has
$\epsilon=0$ leading to an unknown constant factor. This factor
can, in principle, be calculated using the approach presented in
this paper. The resulting expression for $G_e$ reads
\begin{equation}
G_e(0,t)=\frac{\mathrm{const}}{\sqrt{t \ln t}} \left[ 1+\mathcal
O\left(\frac{1}{\ln t} \right)\right].
\end{equation}
The integral (2.6) is evaluated similarly, yielding
\begin{equation}
G_h(0, t)=\frac{\mathrm{const}'}{\sqrt{t \ln t}}
+\frac{\mathrm{const}''}{\sqrt{\ln t}} \rme^{-\rmi t} G^{*}(0,t)
\end{equation}
with the relative correction of the order of $1/\ln t.$


\ack M.B. Zvonarev's work was supported by the Danish Technical
Research Council via the Framework Programme on Superconductivity.
A partial support from the Russian Foundation for Basic Research
under Grant No. 01-01-01045 and from the programme "Mathematical
Methods in Nonlinear Dynamics" of Russian Academy of Sciences are
also acknowledged.


\end{document}